\documentclass[]{spie}  

\usepackage{amsmath,amsfonts,amssymb}
\usepackage{graphicx}
\usepackage[colorlinks=true, allcolors=blue]{hyperref}

\title{FALCO simulations of high-contrast polarimetry with the Nancy Grace Roman Space Telescope Coronagraph Instrument}

\author[a,b]{David S. Doelman}
\author[b]{Hanae Belaouchi}
\author[c]{A.J. Riggs}
\author[c]{Bertrand Mennesson}
\author[b]{Mireille Ouellet}
\author[d]{Rob G. van Holstein}
\author[a]{Jeroen Rietjens}
\author[a]{Henk Hoevers}
\author[b]{Frans Snik}

\affil[a]{SRON Netherlands Institute for Space Research, Niels Bohrweg 4, 2333 CA, Leiden, The Netherlands}
\affil[b]{Leiden Observatory, Leiden University, P.O. Box 9513, 2300 RA Leiden, The Netherlands}
\affil[c]{Jet Propulsion Laboratory, California Institute of Technology, Pasadena, California 91109, USA}
\affil[d]{European Southern Observatory (ESO), Alonso de Córdova 3107, Vitacura, Casilla 19001, Santiago de Chile, Chile}

\authorinfo{Further author information: (Send correspondence to D.S.D)\\D.S.D.: E-mail: D.S.Doelman@SRON.nl}

\begin{document} 
\maketitle

\begin{abstract}
The Coronagraph Instrument of the Nancy Grace Roman Space Telescope (Roman Coronagraph) will be capable of both total intensity and polarization measurements of circumstellar disks. The polarimetric performance is impacted by polarization effects introduced by all mirrors before the Wollaston prisms. In this paper, we aim to characterize these effects for the Roman Coronagraph in bands 1 and 4 using the FALCO and PROPER packages. We simulate the effect of polarization aberrations that impact the polarimetric contrast and the instrumental polarization effects to study the polarimetric accuracy. We include spacecraft rolls, but leave out systematic camera noise. We find that polarimetric differential imaging (PDI) improves the contrast by a factor of six. The PDI contrast of $\sim8\times10^{-11}$ is limited by polarized speckles from instrumental polarization effects and polarization aberrations. By injecting polarized companions with at various contrast levels and demodulating their polarimetric signal, we recover their source Stokes vector within 2\%. 
\end{abstract}

\keywords{Polarimetry, Roman Space Telescope, high-contrast imaging, direct imaging}

\section{INTRODUCTION}
\label{sec:intro}  

Polarimetric high-contrast imaging observations of circumstellar disks and exoplanet atmospheres can yield a wealth of information about their gaseous and dusty constituents.
Measurements of the angle and degree of linear polarization in circumstellar disks give the scattering phase function, which is highly dependent on grain sizes and dust grain morphology \cite{milli2019optical,chen2020multiband}. 
Determination of the scattering phase function through direct imaging polarimetry gives a unique window for characterizing these disks and removes some of the degeneracies from total intensity images. 
Similarly, for exoplanets the scattering phase function is shaped by Rayleigh scattering, yet is strongly impacted by the presence of hazes and clouds \cite{mclean2017polarimetric,stam2004using,rossi2021spectropolarimetry}. 
In addition, polarimetry can improve the signal-to-noise of measurements of circumstellar disks and exoplanets by removing the unpolarized direct star light, revealing the faint signals of polarized scattered light \cite{quanz2011very,stolker2016shadows}.
This makes polarimetry with future space telescopes a powerful tool to study circumstellar disk morphology, planet-disk interactions, and ultimately detecting liquid water on rocky exoplanets \cite{vaughan2023chasing}. 

\noindent The Nancy Grace Roman Space telescope\cite{spergel2015wide} (Roman) is a NASA observatory with a Hubble-sized primary mirror (2.4 m) set to launch before June 2027. 
Roman will mainly focus on the study of dark energy and dark matter with the Wide Field Imager but has a technology demonstrator Coronagraph Instrument. 
The Roman Coronagraph Instrument (Roman/CGI) will be the first dedicated high-contrast imaging instrument with active wavefront control \cite{mennesson2018wfirst,noecker2016coronagraph}.
Simulated performances predict that Roman/CGI will be able to reach extreme contrast ratios ($~10^{-9}$) and small inner-working angles (3$\lambda/D$)\cite{kasdin2020nancy, krist2018wfirst}. 
Although polarimetry is no longer a technology demonstration requirement, Roman/CGI is equipped with two Wollaston prisms oriented at 0 and 45 degrees enabling polarimetry of extended sources at these extreme contrast ratios \cite{groff2021roman}. 
The expected performance of this polarimetric mode is limited by calibration and flat-fielding errors, and the expected error on the linear polarization fraction has been estimated to be at the 3\% level \cite{zellem2022nancy,maier2022flatfield}. 
With careful in-orbit calibration this might be improved.
For example, observations of extended sources like Uranus or Neptune can improve flat-field calibration \cite{maier2022flatfield}. 
Moreover, instrumental polarization effects can be characterized by observing both polarized and unpolized standard stars and extended face-on disks with known polarization (e.g. TW Hya).

\noindent The performance of polarimetric instruments can be characterized by two metrics, the polarimetric sensitivity and the polarimetric accuracy \cite{snik2013astronomical}.
Polarimetric sensitivity of an instrument is given by the smallest polarimetric signal the instrument can detect. 
The polarimetric accuracy is the degree in which the instrument can measure or recover the true polarimetric signal after calibration. 
Accurate and sensitive polarimetry is complicated by instrumental polarization effects. 
Instruments and the telescope can introduce polarization signals and cause unpolarized sources to appear polarized, which is called instrumental polarization. 
Instrumental polarization is caused by diattenuation, where the two orthogonal polarization states have different Fresnel reflection coefficients or different transmissions. 
If there are also changes in the relative phase caused by retardance of optical components, crosstalk between the polarization states is introduced. 
Non-zero retardance can cause changes in the measured angle of linearly polarized light or generate circular polarization, thereby reducing the polarimetric efficiency.
Both instrumental polarization and polarization crosstalk can be modeled through ray-tracing if all coatings are known, or measured through component and/or end-to-end measurements with known input \cite{chipman1995mechanics}.
A system Mueller matrix can be calculated from the individual contributions of all optical elements and this Mueller matrix can be used to recover the true source Stokes vector from observations.
These models, however, might not be good enough to calibrate the instrument down to 1\% polarimetric accuracy. 
Measurements of components or parts of the system, combined with observing polarimetric calibration standards are required to go beyond this level. 
Such instrument calibration has successfully been applied to the polarimetric modes of VLT/SPHERE \cite{de2020polarimetric,van2020polarimetric}, Gemini/GPI \cite{wiktorowicz2014gemini,millar2016gpi}, and SCExAO/CHARIS \cite{van2020calibration,gj2021full}.

\noindent Besides instrumental polarization, a second effect that will impact the performance of especially high-contrast imaging polarimeters are polarization aberrations \cite{chipman1987polarization,mcguire1987polarization}. 
Polarization aberrations are changes in amplitude and/or phase for orthogonal polarization states that create polarization structure in the point-spread function. 
These aberrations are created when the angle of incidence is not constant over the cross-section of a beam of light reflecting off mirrors \cite{chipman2015polarization,van_Holstein_2023}.
This happens with powered optics, like the primary and secondary mirror, or when converging beams are reflected off flat mirrors like tertiary mirrors of Nasmyth telescopes\cite{chipman2015polarization,breckinridge2015polarization}, and has been characterized for a variety of telescopes \cite{anche2018analysis,anche2023estimation,anche2023polarimetric,anche2023polarization}.
The local changes in angle of incidence result in locally varying s- and p-polarization states for the Fresnel coefficients, resulting in locally changing reflection amplitudes and locally varying phases, introducing beam shifts\cite{van_Holstein_2023}.
The detrimental effect on the PSF of the phase aberrations is much stronger than that of the amplitude aberrations.
\cite{van_Holstein_2023}.
We note that these aberrations can not be corrected with deformable mirrors, as these operate on both polarization states in the same way.
The polarization aberrations therefore impact the stellar suppression for all high-contrast imaging systems, not only polarimeters \cite{schmid2018sphere,van2020polarimetric,millar2022polarization}.
However, for high-contrast polarimeters these aberrations, combined with instrumental polarization effects, also impact the polarimetric sensitivity.
The difference in the speckle field between two orthogonal polarization states, creates polarized speckles and spurious signals \cite{sanchez1992instrumental,sanchez1994instrumental}. 
Understanding the sensitivity limits requires modelling of the polarization aberrations with polarization ray-tracing and end-to-end modelling of high-contrast imaging systems in the presence of these aberrations and wavefront control. \\
In this paper, we simulate the performance of polarimetric observations with Roman/CGI in the presence of instrumental polarization effects and polarization aberrations using the FALCO software for optical propagation and wavefront control.
First, we interpret the published Mueller matrix of the Roman/CGI instrument in Section \ref{sec:MM} and construct the Jones pupils using the Mueller matrix as input in Section \ref{sec:Jones}.
Next, we simulate the post-coronagraphic point-spread functions after separation of the linear polarization states with the Wollaston prisms in Section \ref{sec:PSFs}. 
We demodulate these PSFs to reconstruct the source Stokes vector of off-axis companions in Section \ref{sec:Demodulate} and discuss the impact of instrumental polarization effects and polarization aberrations on the polarimetric sensitivity and accuracy in Section \ref{sec:Discussion}. Finally, we draw conclusions in Section \ref{sec:Conclusion}.\\

\section{Understanding the Roman space telescope coronagraph Mueller Matrix}
\label{sec:MM}
\begin{figure} [ht]
   \begin{center}
     \includegraphics[width = \linewidth]{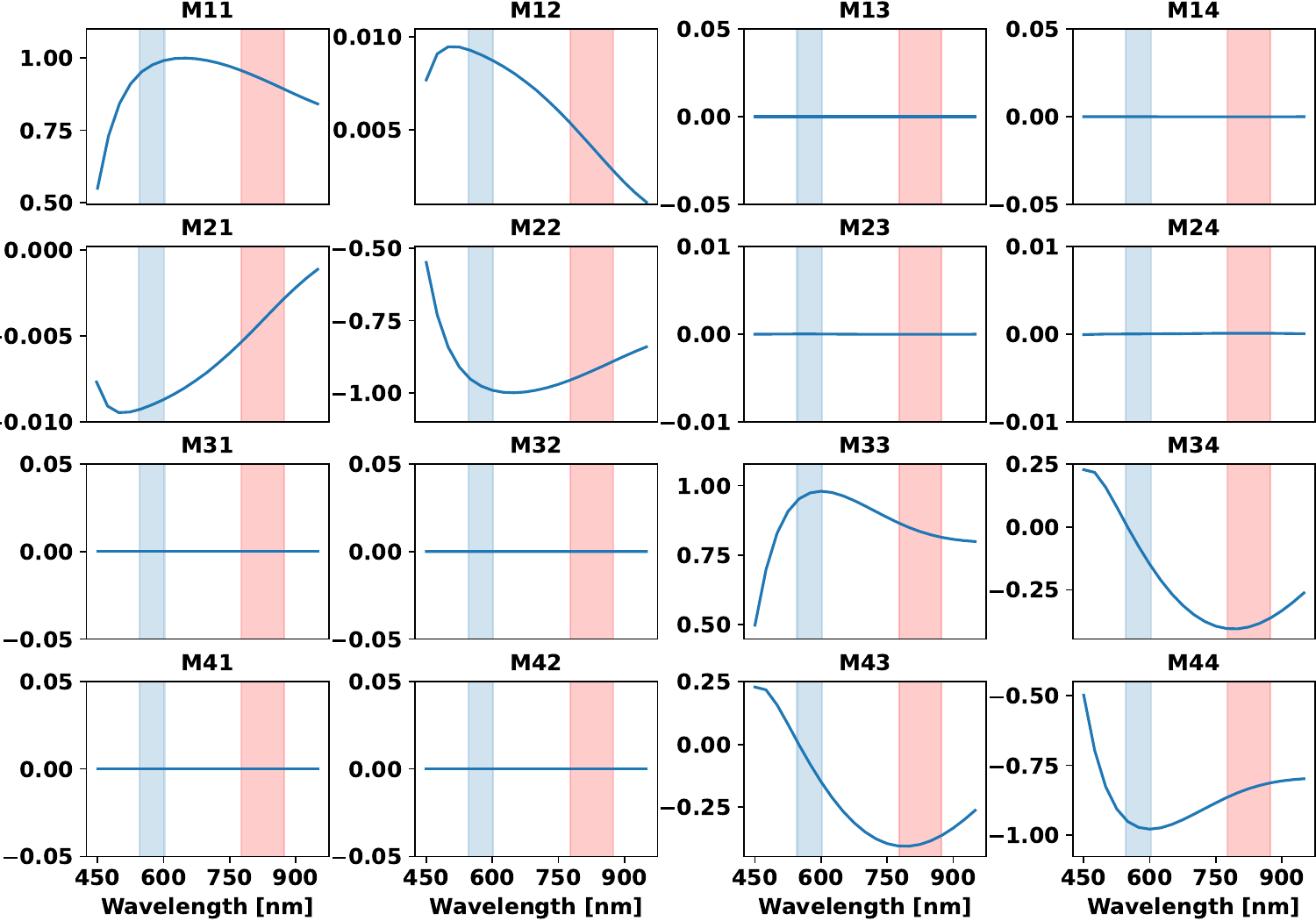}
   \end{center}
   \caption[example] 
   { \label{fig:roman_mm} 
   The published wavelength-dependent Roman/CGI coronagraph Mueller matrix. The two observing bands that have a polarimetric mode, band 1 and band 4, are highlighted in blue and red respectively.}
   \end{figure}

\begin{figure} [ht]
   \begin{center}
     \includegraphics[width = \linewidth]{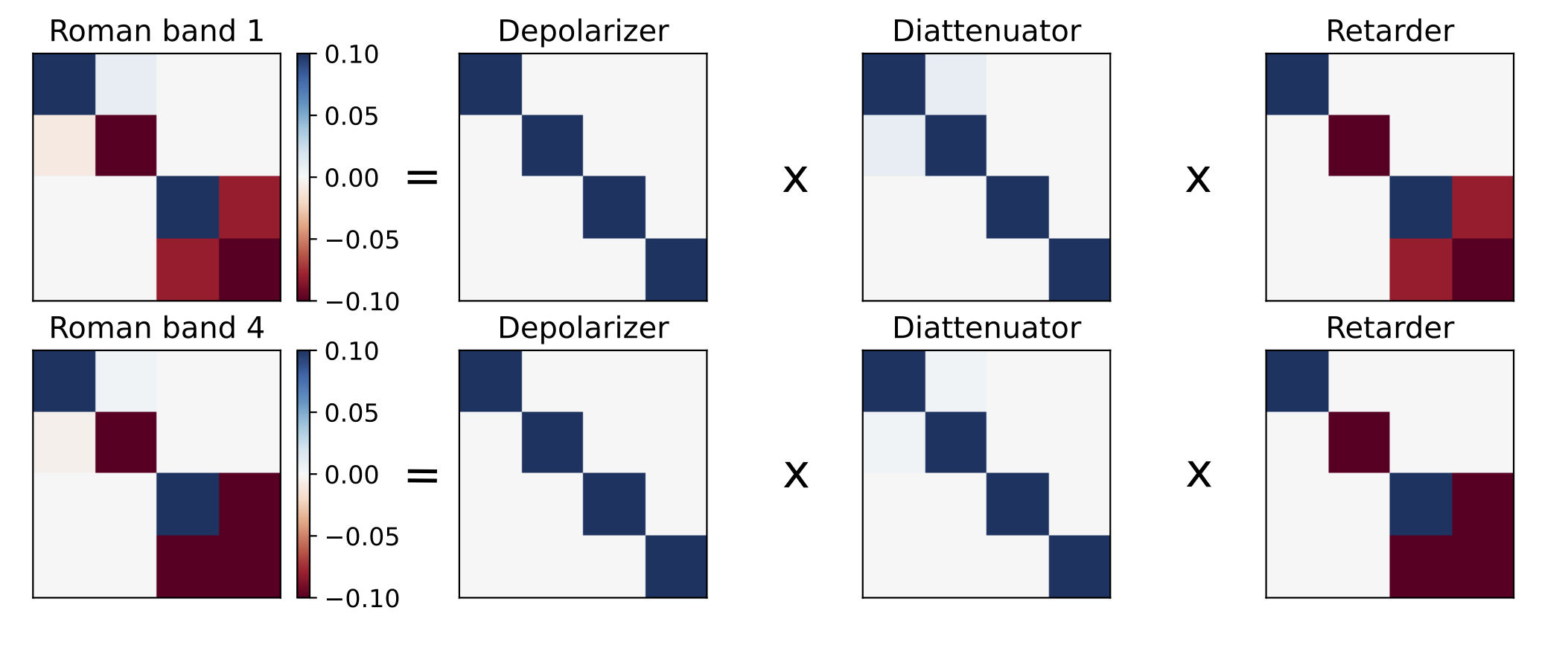}
   \end{center}
   \caption[example] 
   { \label{fig:LC_mm} Lu-Chipman decomposition of the Roman Mueller matrices in band 1 and 4 into a depolarizer, diattenuator, and retarder Mueller matrix. We saturate the colormap to show the retarder and diattenuator components.
   }
   
   \end{figure}

\begin{figure} [ht]
   \begin{center}
   \begin{tabular}{c} 
   \includegraphics[width=\linewidth]{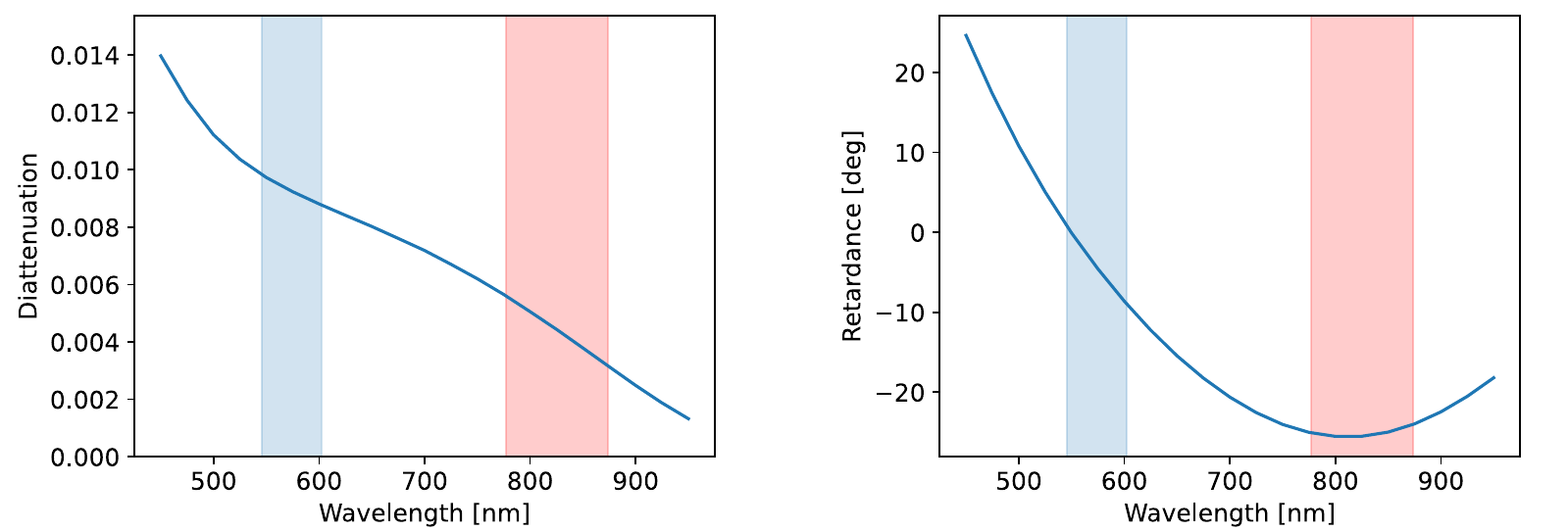}
      \end{tabular}
   \end{center}
   \caption[example] 
   { \label{fig:ret_diat} Diattenuation and retardance as function of wavelength as obtained from the Lu-Chipman decomposition and fitting of the individual matrices.}
   \end{figure}

The Roman Space telescope coronagraph Mueller matrix has been calculated by the Roman Coronagraph project team for 21 wavelengths ranging from 450 nm to 950 nm using end-to-end modeling and has since been released to the science community\footnote{\url{https://roman.ipac.caltech.edu/docs/Roman-Coronagraph-Optical-Model-Mueller-Matrices-450-to-950nm.pdf}}. 
These Mueller matrices describe how the coronagraphic instrument changes the source Stokes vector to an output Stokes vector that is seen by the Wollaston prisms at the end of the CGI instrument.
The wavelength-dependent Mueller matrix is shown in Figure \ref{fig:roman_mm}.
  The Mueller matrix has significant off-axis terms for the M34 and M43 terms, showing the presence of a retarder component and non-zero M12 and M21 components that can come from a diattenuation and/or depolarization component. 
  Furthermore, the diagonal elements swap sign, indicating that there might be coordinate rotations or flips. 
  Our goal is to implement this Mueller matrix in the instrument simulations that use the Jones formalism to propagate the polarization aberrations.
  To this end, we will need to decompose the RST coronagraph Mueller matrix into a few components we can separately turn into its respective Jones matrix.
   We use Lu-Chipman decomposition, a technique that decomposes any Mueller matrix into a retarder Mueller matrix, a diattenuation Mueller Matrix, and a depolarization Mueller matrix. 
   For each wavelength we apply the Lu-Chipman decomposition and retrieve the three Mueller matrices. 
   The Mueller matrices for Band 1 and 4 are displayed in Figure \ref{fig:LC_mm}.
      \\
 We find that the depolarization matrices closely resemble a unity matrix, while the diattenuation matrices show non-zero diattenuation in M12 and M21, which we can extract. 
 The retarder matrix is not as straightforward to interpret as the matrix shows signs of a coordinate rotation and flip with the alternating sign in the diagonal matrix elements.
We fit a rotated retarder and a fold mirror Mueller matrix, a unity matrix with M33 and M44 being -1. 
However, this does not result in a Mueller matrix that fits the retarder matrix for any wavelength.
Instead, we can only generate the retarder matrix when the coordinate system after the retarder stays rotated as a whole, at an angle of 90 degrees. 
The retarder is thus given by the following Mueller matrix:
\begin{equation}
M_{\text{ret, decomposed}} = M_{\mathrm{mirror}} M_{\mathrm{ret}}(\lambda)  M_{\mathrm{rotation}}(90^\circ),
\label{eq:ret}
\end{equation}
where $M_{mirror}$ is the fold mirror matrix described above, $M_{rotation}(90^\circ)$ is a standard rotation matrix for a 90 degree rotation, and $M_{ret}(\lambda)$, is a retarder Mueller matrix.
For each wavelength we fit the retardance and minimize the difference between $M_{ret, decomposed}$ and the Lu-Chipman decomposed matrix.
The retrieved diattenuation and retardance are shown in Figure \ref{fig:ret_diat}.

The diattenuation decreases as function of wavelength monotonically and in a close-to linear fashion, while the retardance is close to zero in band 1 and has a maximum deviation from zero in band 4, reducing the polarimetric efficiency.
We use these values to generate the Jones matrices for the FALCO simulations. This includes the mirror flip and 90 degree rotation as shown in Eq. \ref{eq:ret}.

\section{The Roman coronagraph Jones pupils}
\label{sec:Jones}
To simulate the effect of polarization aberrations, we use the open source
Fast linearized coronagraph optimizer (FALCO) software \cite{riggs2018fast} \footnote{\url{https://github.com/ajeldorado/falco-python}}. 
FALCO comes with models for the full Roman/CGI instrument and these models include polarization aberrations.
FALCO can generate four phase maps for each of the Jones pupil elements. 
The maps are generated for different polarization states by a user input of [-2,-1,1,2], and the coordinate system of the Jones matrix that belong to it are defined in the code which we copied to Table \ref{tab:falco_coord}.
From these definitions we conclude that the FALCO coordinate system includes a pre-computed rotation of $45$ degrees. 
This rotation equalizes the electric field amplitude of all four Jones pupils and the phase is close to zero, except for the few nm aberrations that are introduced, which makes it easy to simulate and compare results. 
 \begin{table}[ht]
\caption{Definition of polarization states in FALCO for the Roman coronagraph simulations. } 
\label{tab:falco_coord}
\begin{center}       
\begin{tabular}{c|c|c} 
Falco input & orientation in & orientation out\\
\hline
-2 &	-45	deg 	& 	Y 	\\
-1&	-45	deg 	& 	X \\
 1&	45	deg 	& 	X 	\\
 2&	45	deg 	& 	Y 	\\
\end{tabular}
\end{center}
\end{table}
The polarization aberration maps are regarded as a weighted sum of the first 21 Zernike modes and the coefficients are precomputed using end-to-end polarization ray-tracing. 	
We note that the average of each of the phase aberration maps is zero, meaning that all instrumental polarization has been removed, and this shows why we analyzed the published Roman Coronagraph Mueller matrix in the previous section.
In addition, the common aberration between all four maps has also been removed and they average to zero. 

We do not attempt to insert some common aberration, as we assume that focal plane wavefront sensing will take most of this out and other type of aberrations are included in the FALCO simulations.\\
Using the definitions in Table \ref{tab:falco_coord}, we define our own Jones pupil matrix based on the four inputs:
\begin{equation}
\mathbf{J}_{\text{FALCO}}=\begin{pmatrix} -1 & 1 \\ -2 &2  \end{pmatrix} = \begin{pmatrix} -45  \rightarrow  X & 45  \rightarrow  X \\ -45  \rightarrow  Y & 45  \rightarrow  Y \end{pmatrix}.
\label{eq:Jones}
\end{equation}

The FALCO Jones pupil and the derotated Jones pupil are presented in Appendix \ref{app:Jones}. Next, we integrate the retardance, diattenuation, mirror and rotation Jones matrices from the Roman coronagraph Mueller matrix in Section \ref{sec:MM}. 
The total instrumental Jones matrix is now given by:
\begin{equation}
J_{\text{Roman}} =  J_{\text{mirror}} J_{\text{ret}}(\lambda) J_{\text{rotation}}(90^\circ)  J_{\text{diat}} J_{\text{FALCO*}} J_{\text{rotation}}(-45^\circ),
\end{equation}
  \begin{figure} [ht]
   \begin{center}
   \begin{tabular}{cc} 
   \includegraphics[width=0.47\linewidth]{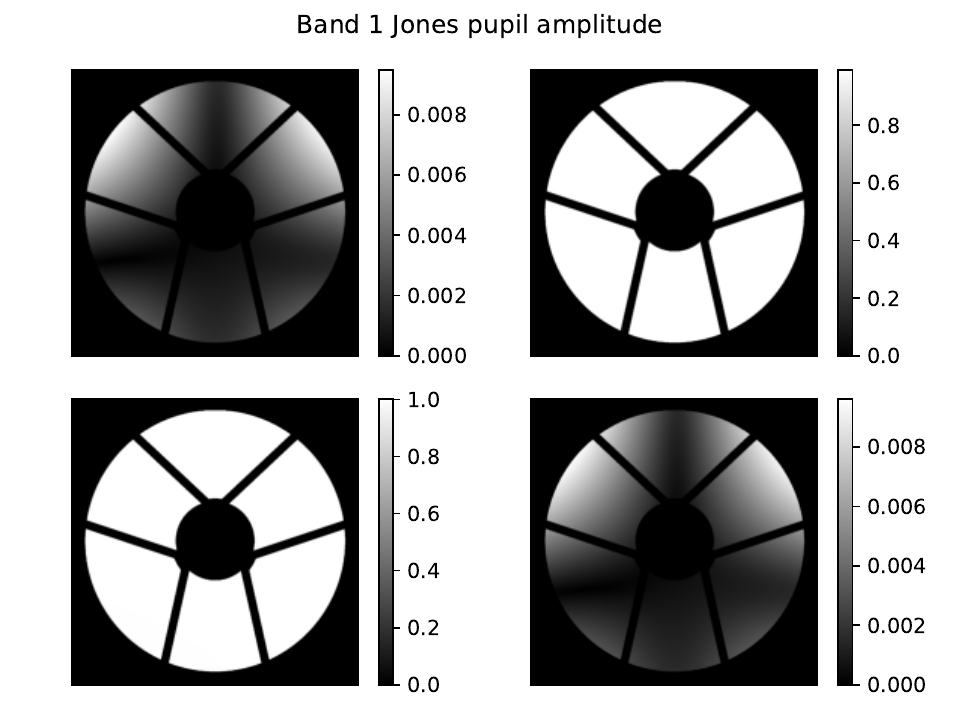} \includegraphics[width=0.47\linewidth]{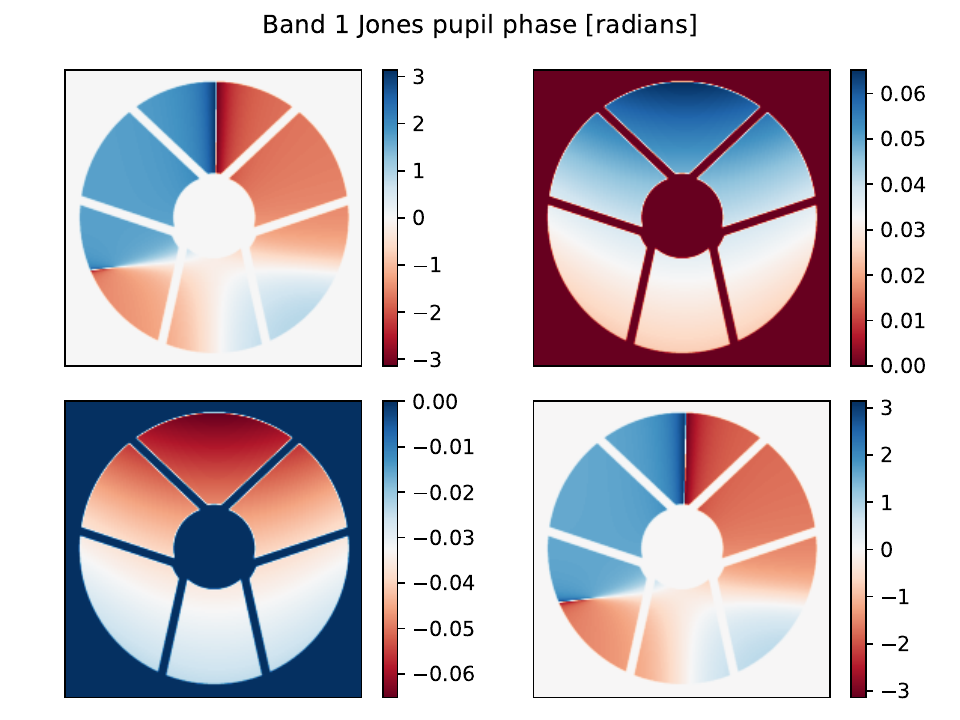}
      \end{tabular}
   \end{center}
   \caption[example] 
   { \label{fig:part_J_b1} The complete Roman Jones pupils of band 1, corresponding to the coordinate system of the published Mueller matrix.}
   \end{figure}

    \begin{figure} [ht]
   \begin{center}
   \begin{tabular}{cc} 
   \includegraphics[width=0.47\linewidth]{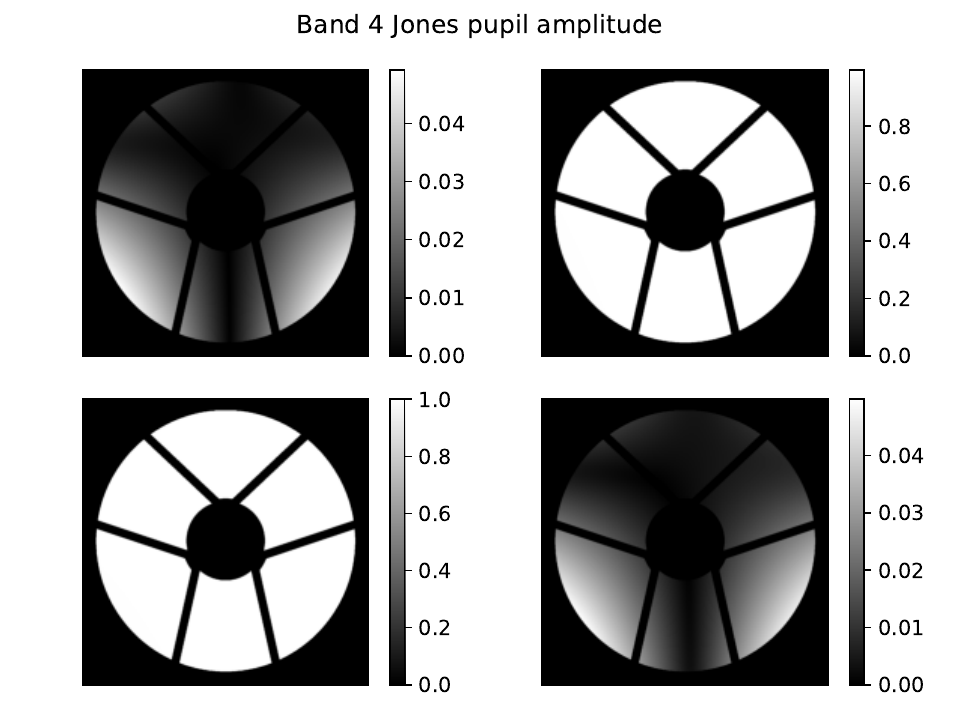} \includegraphics[width=0.47\linewidth]{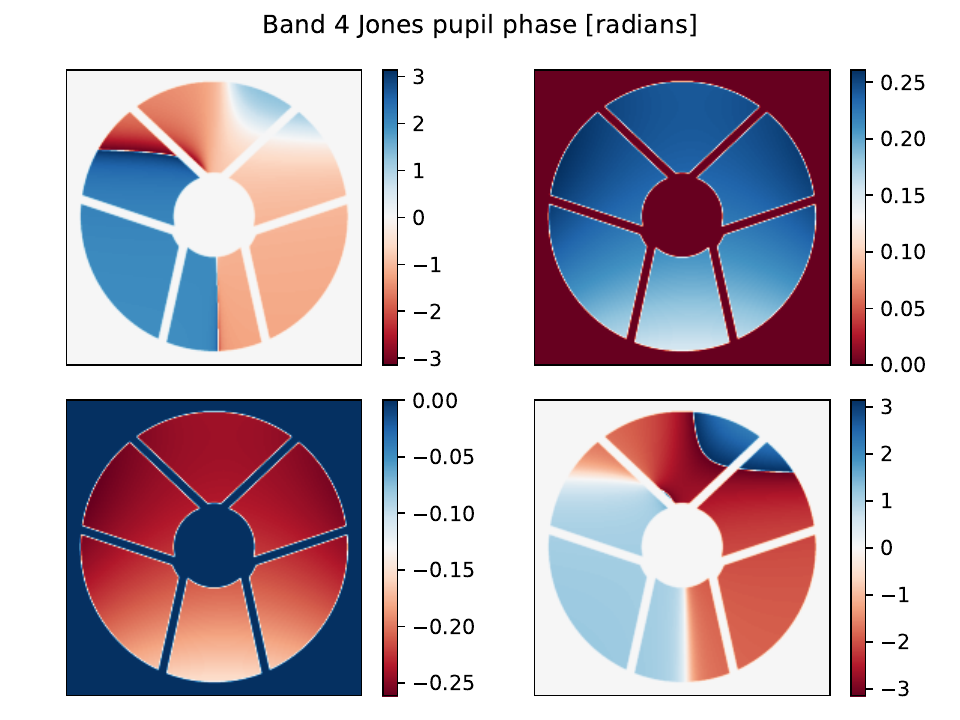}
      \end{tabular}
   \end{center}
   \caption[example] 
   { \label{fig:part_J_b4}  The complete Roman Jones pupils of band 4, corresponding to the coordinate system of the published Mueller matrix.}
   \end{figure} 
where $J_{\text{FALCO*}}$ is the standard FALCO Jones matrix with J12 multiplied by $-1$ and $J_{\text{mirror}}$ is a diagonal 2x2 matrix with (1, -1) as the diagonal terms.
We realize that the many coordinate rotations and flips are not well demonstrated in a figure with coordinate definitions, however, this exercise is only to have the FALCO simulations that use the Jones formalism match with the published Mueller matrix. 
This published Mueller matrices define the coordinate system from the sky to the Wollaston plane, and we confirm that the $J_{\text{Roman}}$ matrix matches the published matrix when averaged over the pupil.

\section{Post-wollaston point-spread functions}
\label{sec:PSFs}
From this point forward we focus on band 4 where the instrumental crosstalk is larger than in band 1, making it more challenging for polarimetry. 
However, the main reason for this choice is that likely due to version control of the multiple python packages we were unable to get a better contrast than $2\times 10^{-8}$ in band 1, while we reached reported literature values for band 4.  
We use FALCO in monochromatic mode to propagate the Jones pupils element-wise through the complete optical system to the science camera and adapt FALCO to save the electric fields at the science camera plane.
We verify that the piston terms for each element are not removed during this propagation and that we can use the focal plane electric fields to construct the system Mueller matrix for the focal-plane.
Constructing the PSF Mueller matrix is required to be able to calculate the PSFs of an unpolarized star or partially polarized sources. 
The low-amplitude and highly aberrated wavefronts of J11 and J22 generate post-coronagraphic PSFs that have poor contrast, so they must be scaled properly to not impact the total coronagraphic performance in simulation.
We normalize the Jones elements by dividing by the sum of the electric field amplitude in the Jones pupil, which in term is normalized to the sum of the electric field amplitude of the unpolarized mode (mode 10).
The resulting Mueller matrix constructed from the normalized Jones matrix is checked by comparing it to the published Roman Mueller matrix and they are consistent. \\
We multiply the focal-plane Mueller matrix by polarizer Mueller matrices rotated by $0^\circ$, $90^\circ$, $-45^\circ$, and $45^\circ$ degrees to get an individual Mueller matrix for the two beams coming out of both Wollaston prisms.
The source stokes vectors can now be multiplied by the four Mueller matrices to obtain Stokes I, i.e the PSF images as seen on the camera. 
We note that this method has been implemented in HCIPy (Por et al. 2018), and we copy this method to the FALCO software.
Adding a spacecraft roll is now relatively straight forward when we stay in the reference frame of the CGI instrument and only requires two steps.
First, we rotate the input stokes vector of all sources by the roll angle. 
Second, we shift any off-axis sources around the optical axis to the rotated position because the PSF from off axis-sources do not rotate in the instrument reference frame. \\
   \begin{figure} []
   \begin{center}
   \begin{tabular}{c} 
   \includegraphics[width=\linewidth]{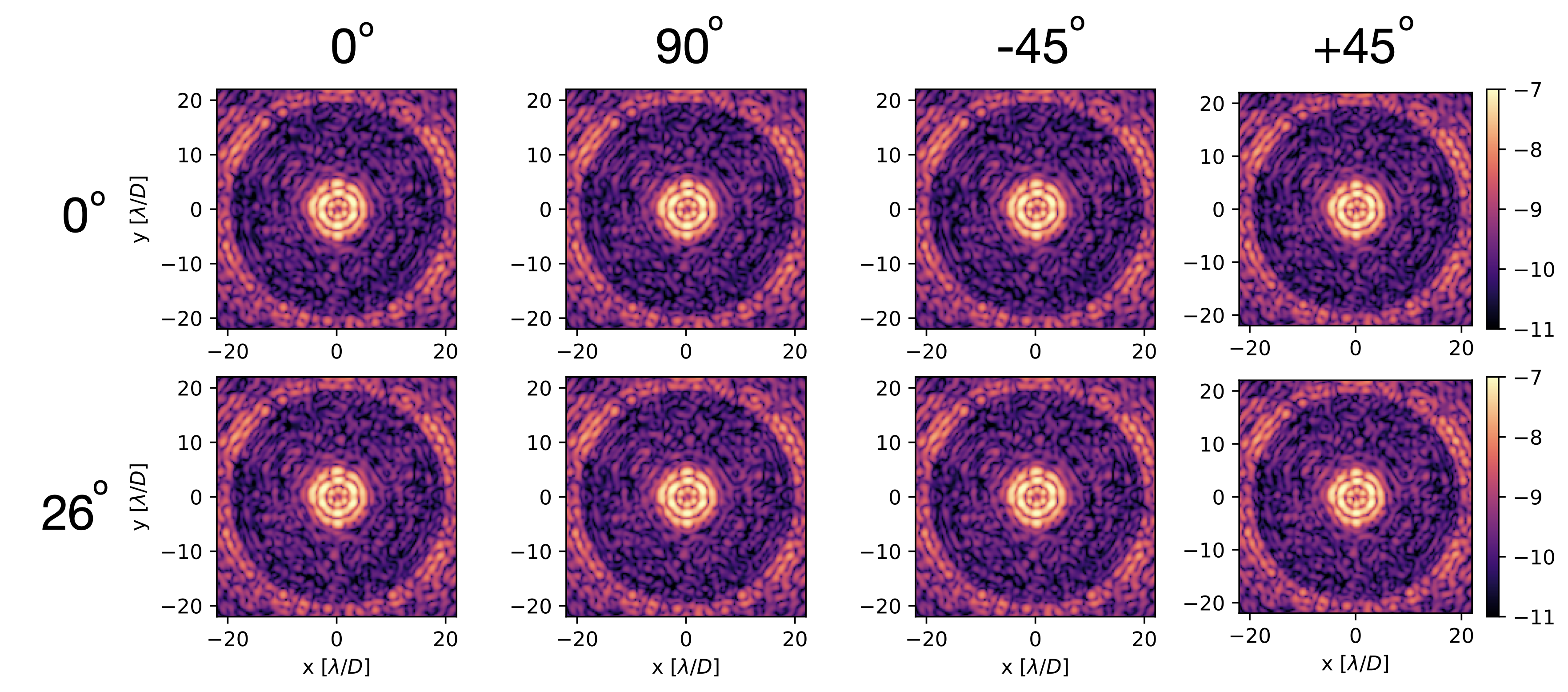} 
         \end{tabular}
   \end{center}
   \caption[example] 
   { \label{fig:woll_images} The band 4 Roman coronagraphic PSFs for an unpolarized star as seen through both Wollaston prisms and at two roll angles, resulting in eight images.  }
   \end{figure} 

\begin{figure} [ht]
   \begin{center}
   \begin{tabular}{c} 
   \includegraphics[width=0.8\linewidth]{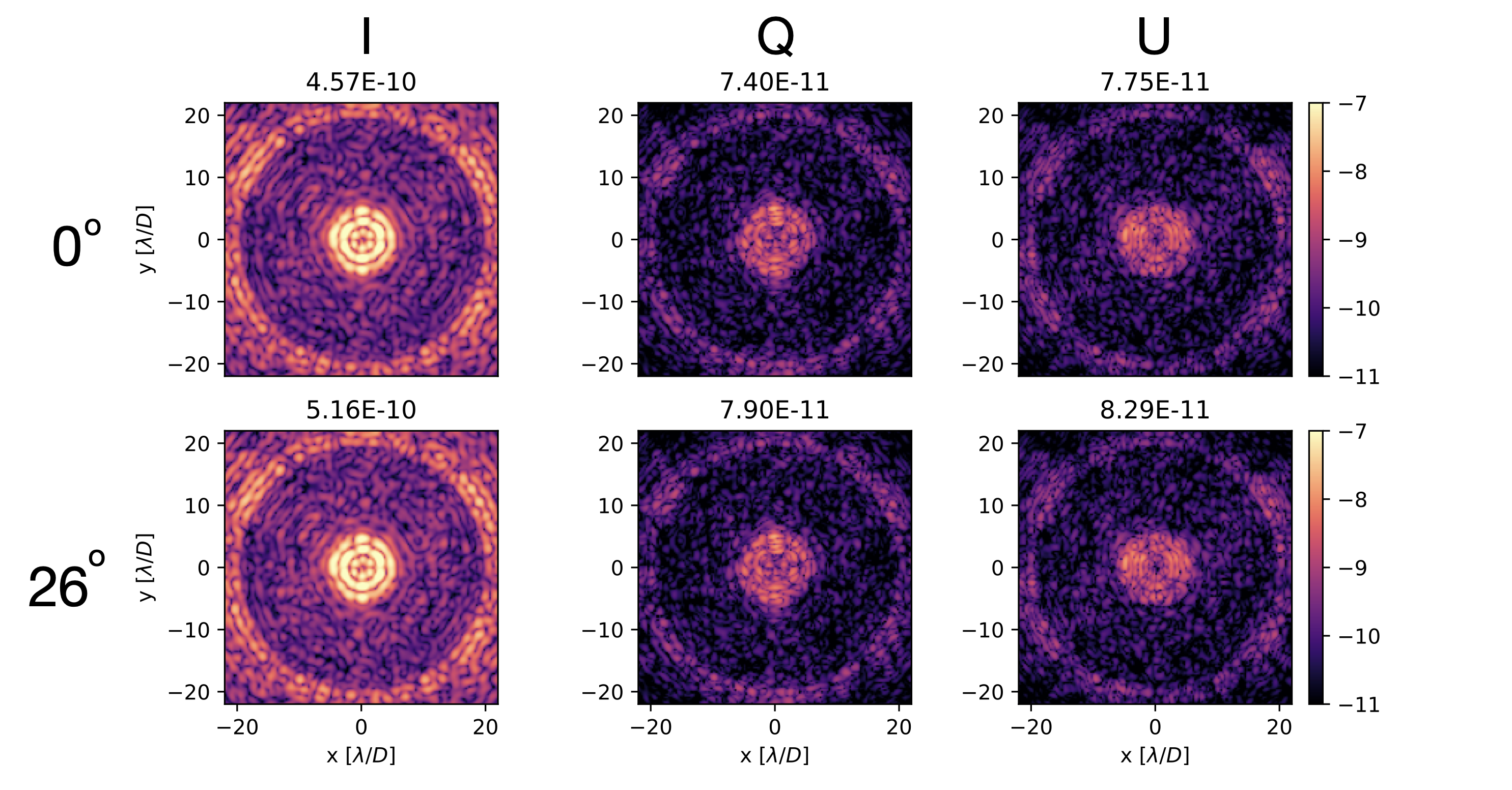} 
         \end{tabular}
   \end{center}
   \caption[example] 
   { \label{fig:pdi}  The total intensity (left) and polarimetric difference images for the $|0^\circ - 90^\circ|$ (center) and $| (-45^\circ) - (45^\circ)|$ (right) band 4 Wollaston images on logarithmic scale. The given contrast level is the mean of the absolute value in the dark hole.}
   \end{figure}

For the wavefront control we first run FALCO in unpolarized mode, mode 10, for 14 iterations and adapt FALCO to save the deformable mirror (DM) commands.
We reach a contrast of $3.8\times10^{-10}$ without changing anything in the wavefront control code.
Next, we apply these DM commands to each of the Jones pupil elements and calculate the four images created by the two Wollaston prisms for two roll angles of 0 and $26^\circ$.
To attempt to introduce a bit of diversity between the two roll angles, we use the 10th iteration wavefront control loop and the 13th iteration of the loop for 0 and $26^\circ$ respectivily.
The resulting images can be seen in Figure \ref{fig:woll_images}, showing that the speckle field is a bit different for all images because of the presence of polarization aberrations.
Next, we add all images per roll angle for our stokes 'I' measurement, and subtract the opposite images for each Wollaston prism to get 'Q' and 'U'. 
These Stokes vectors are in the reference frame of the camera and not in the sky frame, and will be demodulated in Section \ref{sec:Demodulate}.
However, 'Q' and 'U' in the camera frame of reference are still useful as a polarimetric differential images, see Figure \ref{fig:pdi}.
      \begin{figure} [ht]
   \begin{center}
   \begin{tabular}{c} 
   \includegraphics[width=\linewidth]{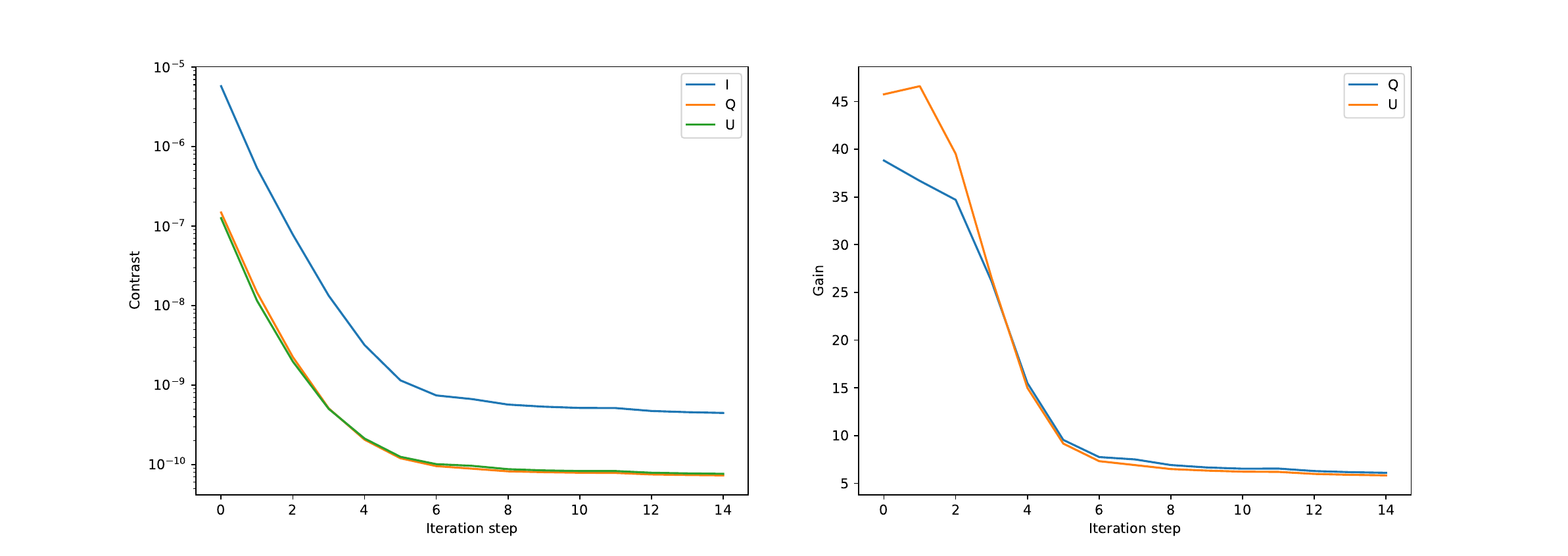} 
         \end{tabular}
   \end{center}
   \caption[example] 
   { \label{fig:con_vs_loop} \textit{Left}: Contrast for the intensity image and two the two 'Q' and 'U' polarimetric differential images in the camera reference frame. \textit{Right:} Gain in contrast for 'Q' and 'U'. }
   \end{figure} 
There is a significant contrast improvement of a factor of six in the polarimetric differential images, meaning that in this idealized simulation a companion with a degree of linear polarization that is larger than 13\% will have an increased signal-to-noise. 
The gain is similar to the reported FALCO polarimetric simulations $|X-Y|$ that use the standard [-2,-1,1,2] set-up \footnote{\url{https://roman.ipac.caltech.edu/docs/WFOV_Polarized_Dataset_Documentation.pdf}}. 
This is expected as the strength of the polarization aberrations has not changed, we mostly focused on adding instrumental polarization from the Mueller matrix. 
We do not take into account additional noise sources like camera noise, flat-fielding errors, and a realistic observing sequence, so the factor six might be an optimistic estimate.
It is still interesting to see how the gain in contrast changes with control loop iteration, as the relative contribution of the polarization aberrations increases with improved contrast.
We simulate the eight Wollaston images and their resulting PDI images for the 14 iterations of the control loop and plot the 'I', 'Q', and 'U' contrast in Figure \ref{fig:con_vs_loop}. 
The gain in contrast is much higher, up to a factor of 40, for the first few iterations and rapidly drops to the factor of six once the dark hole is established.  
It is likely that a cross-talk term between the other aberrations and the polarization aberrations limits the gain for low contrast. 
Expectedly, polarimetric differential imaging does not yield a large difference in contrast between 'Q' and 'U'. 
The correlation between the contrast gain and polarization aberration strength is left for future work.
For now, the next step is to inject planets to verify the contrast gain and retrieve their Stokes vectors in the 'sky' coordinates.

\section{Band 4 exoplanet polarimetry through demodulation}
\label{sec:Demodulate}

We inject planets using the FALCO code that generates off-axis sources for throughput calculation. 
The off-axis sources of the 'compact' mode includes the aberrated Jones Pupil, which we then use to calculate the planet instrumental Mueller matrix.
By multiplying the planet instrumental Mueller matrix with the Wollaston Mueller matrices and the input Stokes vector, we retrieve the planet images without any starlight present.
As explained in the previous section, the spacecraft roll entails a shift in the off-axis PSF and a rotation of the input Stokes vector.
We then define a contrast ratio and co-add the star and planet images.

   \begin{figure} [ht]
   \begin{center}
   \begin{tabular}{c} 
   \includegraphics[width=0.8\linewidth]{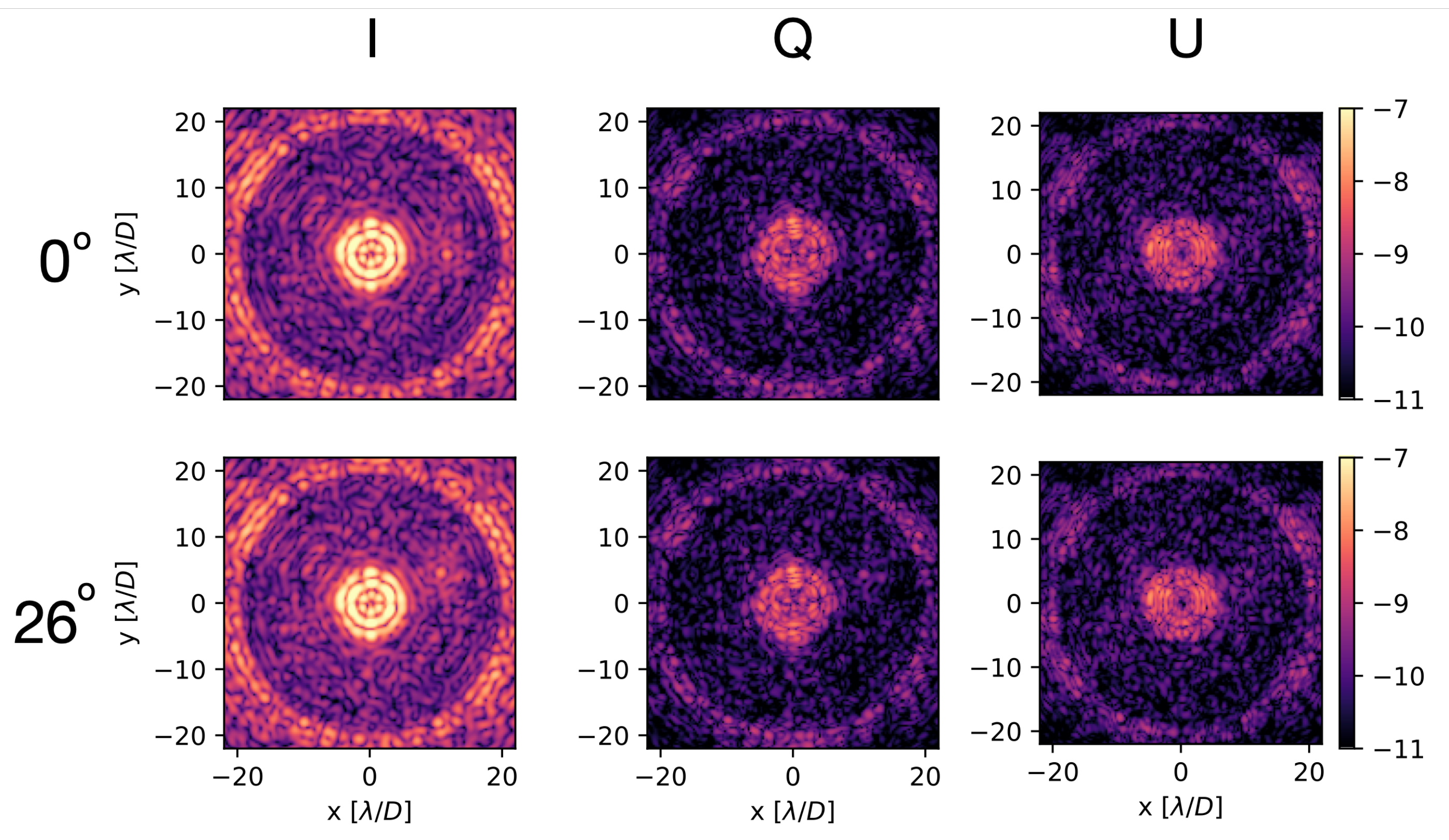} 
         \end{tabular}
   \end{center}
   \caption[example] 
   { \label{fig:pdi_wp} Same as Figure \ref{fig:pdi} with an additional planet with $1\times10^{-8}$ planet with 10\% Q polarization.}
   \end{figure} 

The result is shown in Figure \ref{fig:pdi_wp} for a $1\times10^{-8}$ planet with 10\% Q polarization in the sky frame, i.e. the input Stokes vector is (1,0.1,0,0).
The planet is clearly visible in 'I' and 'Q' for a zero degree roll angle, and all three components for the 26 degree roll angle. 
We confirm that the spacecraft roll is implemented correctly through aperture photometry of the planet. 
The next step is to demodulate the signal into the 'sky' coordinate system and explore the accuracy of the reconstruction.\\
The observing sequence with a spacecraft roll is not different from angular differential imaging for ground-based telescopes. 
Therefore, we can use methods developed for GPI to demodulate the data and find the sky Stokes vector.
We use their least-squares approach to sequence combination, as described in Perrin et al. 2015\cite{perrin2015polarimetry} and van Holstein et al. 2020 \cite{van2020polarimetric}. 
This approach uses that the source Stokes vector is linearly modulated by the system Mueller matrix and for every Wollaston image and roll angle you probe the source Stokes vector in a different way.
If the system Mueller matrix is well characterized, a least-squares fit of the images can be used to retrieve the source Stokes vector.
We calculate the first four elements of the system Mueller matrix $m_{11} \hdots m_{14}$ by multiplying the published Roman CGI Mueller matrix by a (rotated) linear polarizer. 
We vary the polarizer angle to $0^\circ$, $90^\circ$, $-45^\circ$, and $45^\circ$ degrees, and the roll angle by $0^\circ$, $26^\circ$ and retrieve the following linear system:
\begin{equation}
 \begin{pmatrix} I_1 \\ I_2\\ \vdots \\ I_8 \end{pmatrix} =\begin{pmatrix}
  m_{11}(0^\circ,0^\circ) & m_{12} (0^\circ,0^\circ) & m_{13} (0^\circ,0^\circ) &m_{14} (0^\circ,0^\circ)\\
  m_{11}(90^\circ,0^\circ) & m_{12} (90^\circ,0^\circ) & m_{13} (90^\circ,0^\circ) &m_{14} (90^\circ,0^\circ)\\
  m_{11}(-45^\circ,0^\circ) & m_{12} (-45^\circ,0^\circ) & m_{13} (-45^\circ,0^\circ) &m_{14} (-45^\circ,0^\circ)\\
    m_{11}(45^\circ,0^\circ) & m_{12} (45^\circ,0^\circ) & m_{13} (45^\circ,0^\circ) &m_{14} (45^\circ,0^\circ)\\
  m_{11}(0^\circ,26^\circ) & m_{12} (0^\circ,26^\circ) & m_{13} (0^\circ,26^\circ) &m_{14} (0^\circ,26^\circ)\\
  m_{11}(90^\circ,26^\circ) & m_{12} (90^\circ,26^\circ) & m_{13} (90^\circ,26^\circ) &m_{14} (90^\circ,26^\circ)\\
  m_{11}(-45^\circ,26^\circ) & m_{12} (-45^\circ,26^\circ) & m_{13} (-45^\circ,26^\circ) &m_{14} (-45^\circ,26^\circ)\\
    m_{11}(45^\circ,26^\circ) & m_{12} (45^\circ,26^\circ) & m_{13} (45^\circ,26^\circ) &m_{14} (45^\circ,26^\circ)\end{pmatrix}
    \begin{pmatrix}I \\ Q \\ U \\ V \end{pmatrix}.
\end{equation}
 
\begin{figure} [ht]
   \begin{center}
   \begin{tabular}{c} 
   \includegraphics[width=\linewidth]{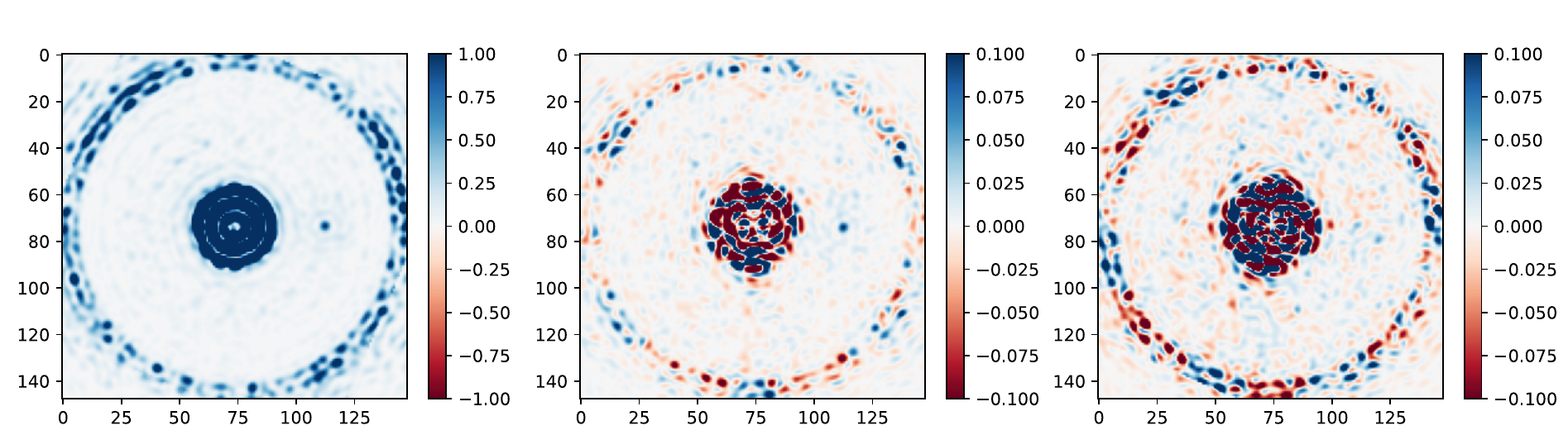} 
         \end{tabular}
   \end{center}
   \caption[example] 
   { \label{fig:plan_demod1} The demodulated I, Q, and U images for a planet with a contrast of $1\times10^{-8}$ and 10\% Q polarization. The images are linearly scaled and normalized on the maximum flux of the planet PSF.  }
   \end{figure}

Following  \cite{perrin2015polarimetry}, we write the compact form of this equation as 
 
 \begin{equation}
I' = M' S,
\end{equation}
which contains three sets of equations for I,Q,U respectively and the least-squares solution is found by the pseudo-inverse of this equation
 \begin{equation}
S_{est} = (M'^T M)^{-1} M'^T I'.
\end{equation}
The images with a spacecraft roll are now derotated so that the observed field is overlapping.
By multiplying the eight (derotated) Wollaston images by the inverse of this demodulation matrix we retrieve the sky Stokes vector for every pixel in the field of view.
We apply this technique to the data with the planet with a contrast of $1\times10^{-8}$ with 10\% Q polarization as shown in Figure \ref{fig:pdi_wp}, and retrieve the estimated source I, Q, and U.
This is shown in Figure \ref{fig:plan_demod1}. 
The estimated Stokes vector = $(1 , 0.1 , 0.02 , 0)$, correctly reconstructing Q at 10\% and a small error in U of 2\%.
The structures and variation in the images of Q and U are the result of different speckle intensities in the eight images used for demodulation.
We observe that the speckle field has a lot of structure in Q and U, however the overall intensity is quite homogeneous as function of separation.

   \begin{figure} []
   \begin{center}
   \begin{tabular}{c} 
   \includegraphics[width=\linewidth]{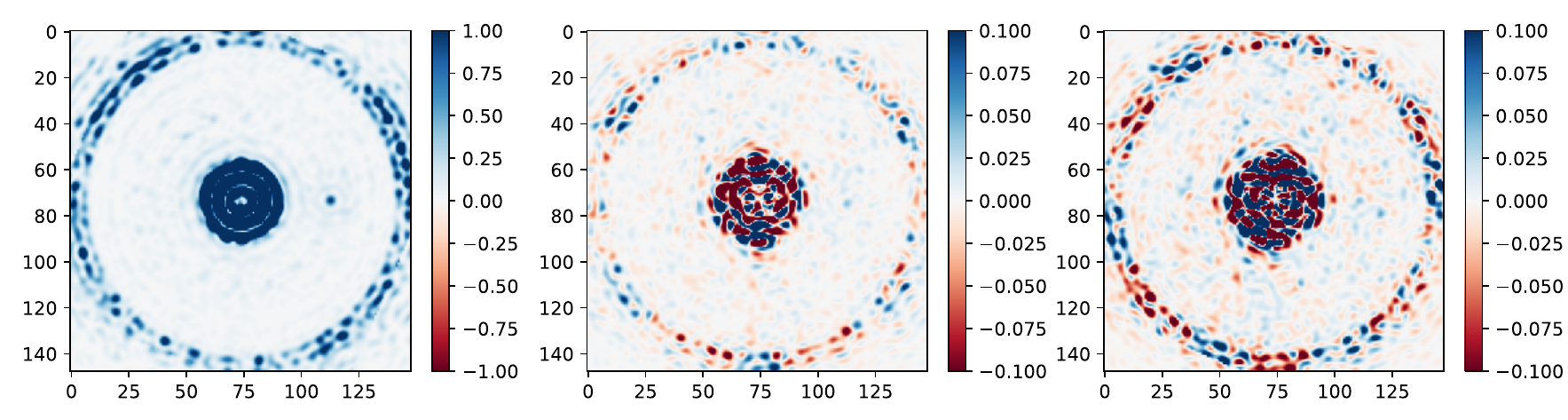} 
         \end{tabular}
   \end{center}
   \caption[example] 
   { \label{fig:plan_demod2} The demodulated I, Q, and U images for a planet with a contrast of $1\times10^{-8}$ and 3\% Q polarization. }
   \end{figure} 

   \begin{figure} []
   \begin{center}
   \begin{tabular}{c} 
   \includegraphics[width=\linewidth]{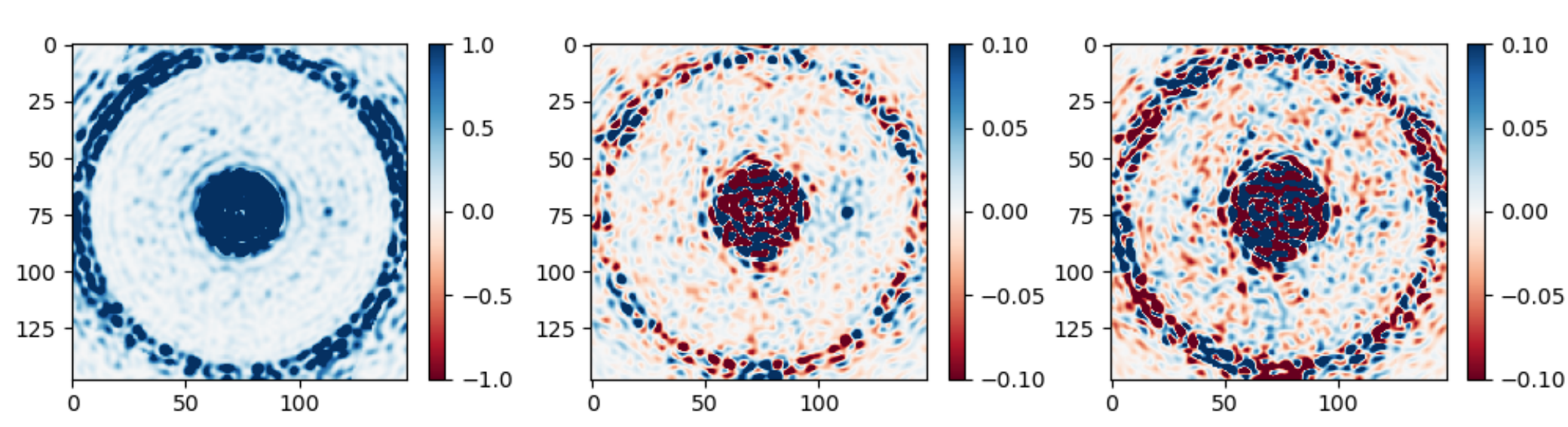} 
         \end{tabular}
   \end{center}
   \caption[example] 
   { \label{fig:plan_demod3} The demodulated I, Q, and U images for a planet with a contrast of $3\times10^{-9}$ and 40\% Q polarization.  }
   \end{figure} 
These speckles are a direct result of the polarization aberrations which not only limits the contrast, but also the polarimetric accuracy. 
We explore this parameter space a bit more, by simulating two more planets with a contrast of $1\times10^{-8}$ and 3\% Q polarization and a planet with a contrast of $3\times10^{-9}$ and 40\% Q polarization. 
The first object has a degree of linear polarization akin a Jovian planet with a tropospheric cloud layer and a stratospheric haze layer, reducing the polarization fraction to a few percent level.
The second object has no cloud or haze layer, and the reflection at quadrature is dominated by Rayleigh scattering resulting in large polarization fractions. 
Their demodulated images are shown in Figure \ref{fig:plan_demod2} and Figure \ref{fig:plan_demod3}. 
We retrieve a Stokes vector of (1, 0.04, 0.01, 0) for the first planet and a Stokes vector of (1, 0.31, 0.0, 0) for the second. 
While the first planet is more challenging as the polarized contrast, the contrast times polarization fraction, is lower, we have a larger reconstruction error for the second planet.
This is likely caused by the normalization of the stokes vector with the planet flux, as the planet intensity is close to the speckle field intensity. 
Overall, the reconstruction is performing well, and seems accurate within 2\% for linearly polarized planets with a contrast around $1\times10^{-8}$. 
We note that these simulations do not include many effects that can worsen the reconstruction, which we will discuss in the next section. 

\section{Discussion}
\label{sec:Discussion}
\begin{figure} [ht]
   \begin{center}
   \begin{tabular}{c} 
   \includegraphics[width=0.8\linewidth]{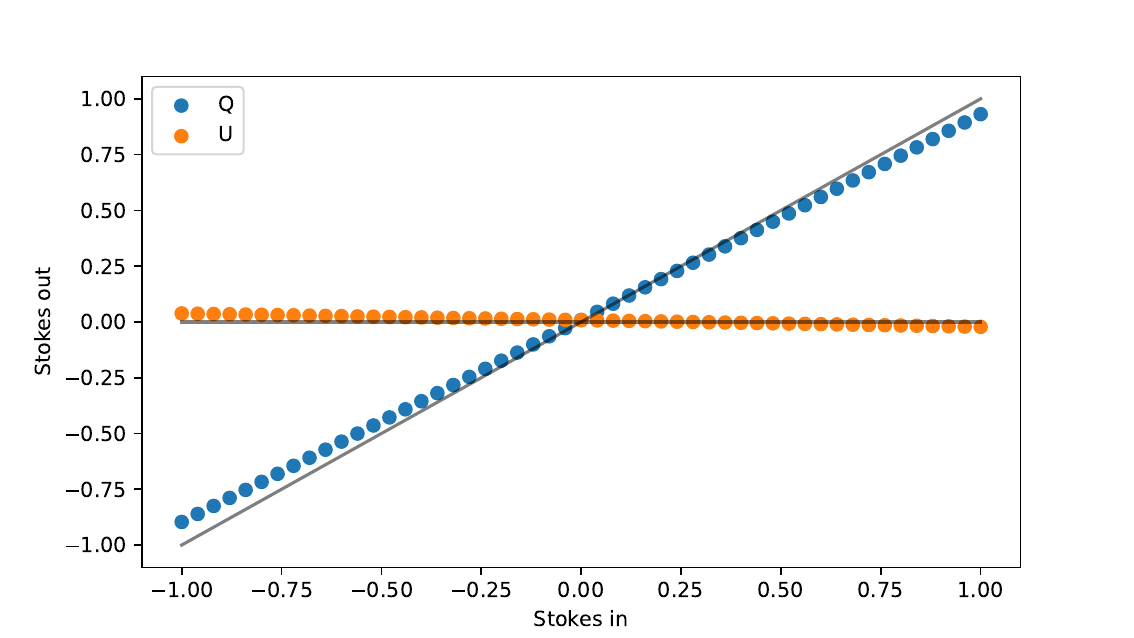} 
         \end{tabular}
   \end{center}
   \caption[example] 
   { \label{fig:stokes_comp} Input vs demodulated Q and U for planets with a contrast of $1\times10^{-8}$ with Q polarization between -1 and 1.}
   \end{figure}

In this paper, we present the methodology to add polarimetry to FALCO based on the published Roman/CGI Mueller matrix and the current implementation of polarization aberrations in FALCO.
As we focus on the methodology, we did not aim to include the most realistic observing sequence and additional noise factors.
The simulation packages for more realistic observing sequences exist, i.e. CGIsim, in addition to packages for realistic camera noise like EMCCD Detect. 
For example, bad pixels have been limiting the performance for VLT/SPHERE \cite{van2021survey}.
The methodology presented in this paper can be easily combined with these packages as the FALCO back-end is the same.
Combined with realistic disk models from MCFOST, we find that all ingredients for a more thorough understanding of the limits of the Roman/CGI polarimetric mode are present.
We leave this work for future papers.\\
The FALCO setup we used for the band 4 simulations were monochromatic with perfect electric field estimation, without realistic drifts in wavefront, jitter, any realistic camera noise. 
This limits the interpretation of the simulations and the associated polarimetric sensitivity and polarimetric accuracy.
An important effect that will likely be a significant contributing factor is the pixel-to-pixel gain variations on the percent level that are corrected with a flat-field correction.
As polarimetric differential imaging relies heavily on subtracting high fluxes with low polarized signal, these gain variations are quickly at the same level as the polarized flux.
The demodulation method described used to retrieve the sky Stokes vector is not robust against these errors. 
An improved flat-field calibration is possible by observing extended sources like Uranus and Neptune.
Another contributing factor will be deviations in the Roman Mueller matrix from the ray-traced simulated one. 
Even in-orbit changes of this Mueller matrix could occur due to degradation of optics or dynamical changes in the telescope or instrument. 
The execution of the calibration campaign of the Roman Space Telescope, both on the ground and in-orbit, will likely determine the actual polarimetric sensitivity and accuracy. 
Observing (extended) sources with known polarization will be the only method for accurate calibration. \\
Despite these considerations, we can still interpret the impact of the polarization aberrations and instrumental polarization assuming it to be towards the optimistic side.
We find that the polarization aberrations do limit the polarimetric sensitivity and the polarimetric accuracy, as the demodulated Stokes vectors have errors on the order of 2\% for the $1\times10^{-8}$ contrast planet.
We look deeper into these limits by simulating planets with a contrast of $1\times10^{-8}$ with Q polarization between -1 and 1 to study the demodulated Stokes vectors. 
The results are shown in Figure \ref{fig:stokes_comp}.
We find that the demodulation for both Q and U is extremely linear and well behaved, yet has a minor residual systematic error. 
Both lines have an offset of 0.01 (1\%) for zero input, the demodulated Q has values between [$-0.89$, $0.93$] and the demodulated U has values between [$-0.04$, $0.02$].
These offsets and maxima and minima are influenced by the speckle field from the polarization aberrations and instrumental polarization effects.
We note again that these results will be on the optimistic side, yet nonetheless they paint a promising picture for Roman polarimetry. \\
For future work we will work on the following things, also in the context of NASA's Community Participation Program:
\begin{itemize}
  \setlength\itemsep{0.2em}
\item Accurate error analysis and error budgeting by including additional noise sources, e.g. flat-fielding errors \cite{maier2022flatfield}.
\item Simulate calibration/validation procedures\cite{zellem2022nancy}
\item Investigate new wavefront control schemes \cite{mendillo2021dual}
\item Include more advanced data-analysis approaches based on ADI and RDI \cite{van2020polarimetric}
\item Target selection
\item Ground-support precursor observations
\end{itemize}

\section{Conclusion}
\label{sec:Conclusion}
\begin{itemize}
  \setlength\itemsep{0.2em}
	\item We present the methodology to add polarimetry to FALCO based on the published Roman/CGI Mueller matrix and the current implementation of polarization aberrations in FALCO.
	\item Decomposing the published Roman Mueller matrix yields the wavelength-dependent retardance of Roman/CGI, which varies between +23 and -23 degrees and is close to zero in band 1.  
	\item The adapted Jones pupil from FALCO is consistent with Jones pupils simulated for other telescopes after modification of the J12 Jones element and a derotation of 45 degrees.
	\item The demodulated Stokes vectors are accurate within 2\% for planets with a contrast of $1\times10^{-8}$, which is an optimistic estimation for the actual performance given the exclusion of realistic noise effects.
	\item Demodulation of eight PSF images from both Wollaston prisms for two spacecraft roll angles enables more accurate de reconstruction. 
	\item The polarimetric contrast is six times better than the total intensity contrast, showing that polarimetric differential imaging can be effective for Roman/CGI for planets with degree of linear polarization of 13\%.
\end{itemize}

\newpage
\appendix
\section{Updating the FALCO Jones pupils}
\label{app:Jones}
\begin{figure} [ht]
   \begin{center}
   \begin{tabular}{cc} 
   \includegraphics[width=0.47\linewidth]{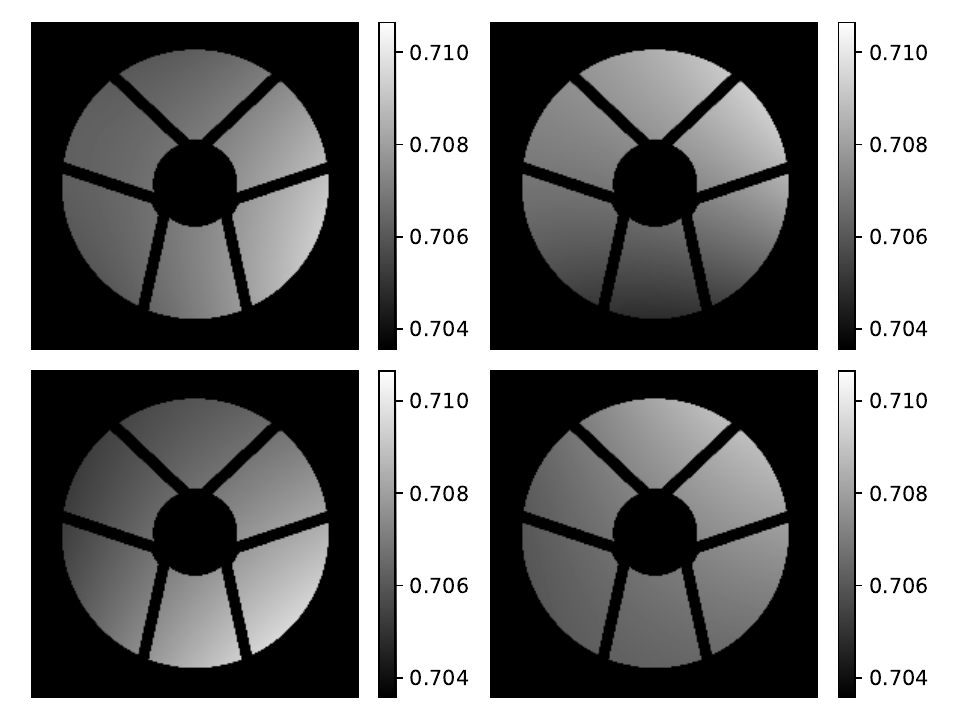} \includegraphics[width=0.47\linewidth]{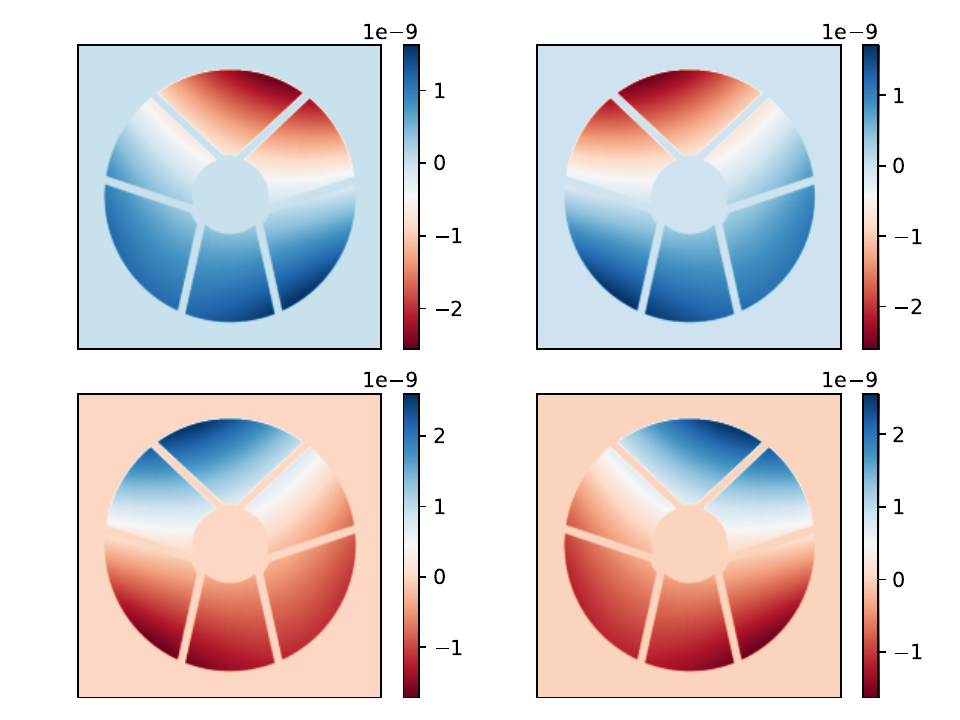}
      \end{tabular}
   \end{center}
   \caption[example] 
   { \label{fig:falco_jpup} The Roman Jones pupils of band 1, showing the amplitude of the electric field on the left and the polarization aberration phase in nanometers on the right. }
   \end{figure}

The FALCO Jones pupil of band 1, as defined in Eq. \ref{eq:Jones}, is shown in Figure \ref{fig:falco_jpup}.
We see that the polarization aberrations in the phase are on the order of $\pm$ 2 nm and the amplitude variations are less than a percent. 
When we define the Jones matrix this way, we find that the electric field amplitudes generate a Jones matrix of a polarizer oriented at 45 degrees. 
This can be understood when taking the dot product of (1,0), and (0,1) with this matrix, both resulting in (0.707,0.707). 
We therefore multiply J12 with $-1$, resulting in a $\pi$ absolute phase shift and creating a Jones matrix that resembles a rotation matrix of 45 degrees. 
A second step towards polarimetry with this Jones pupil is to derotate it to the XY $\rightarrow$ XY coordinate system by derotating the matrix by 45 degrees.
These matrices are more easily interpreted and can then be used to match the Roman Mueller matrix.

We show the band 1 and band 4 Mueller matrices that are derotated and with the J12 sign changed in Figure \ref{fig:part_J_b1} and Figure \ref{fig:part_J_b4}.
The derotated and corrected Roman Jones pupils show a close-to unit matrix in the amplitude and phase in the XY $\rightarrow$ XY coordinate system.
   \begin{figure} [ht]
   \begin{center}
   \begin{tabular}{cc} 
   \includegraphics[width=0.47\linewidth]{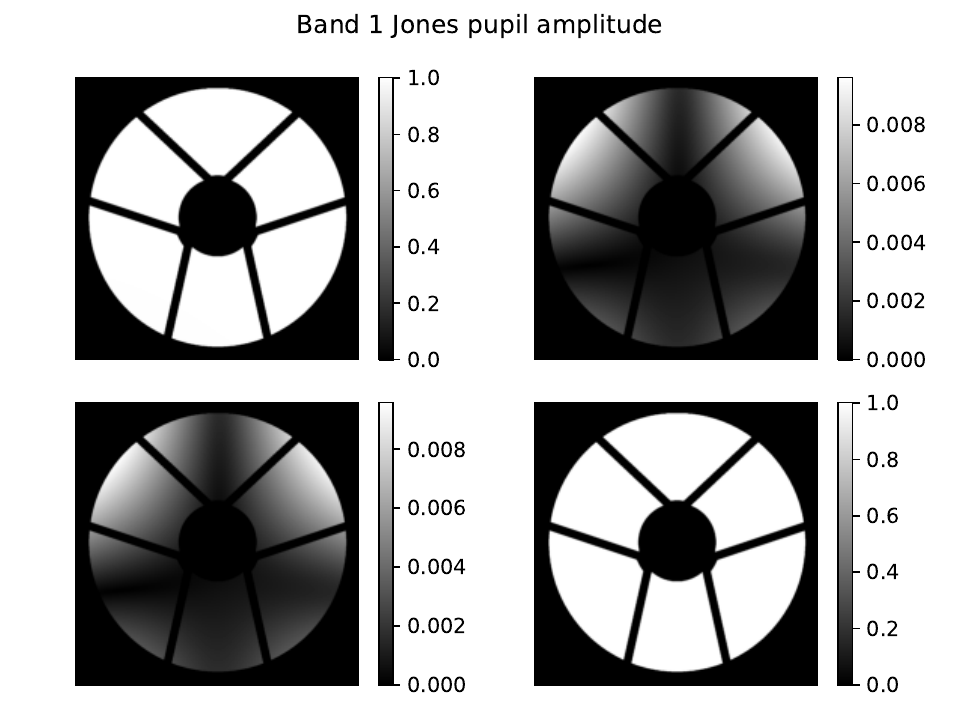} \includegraphics[width=0.47\linewidth]{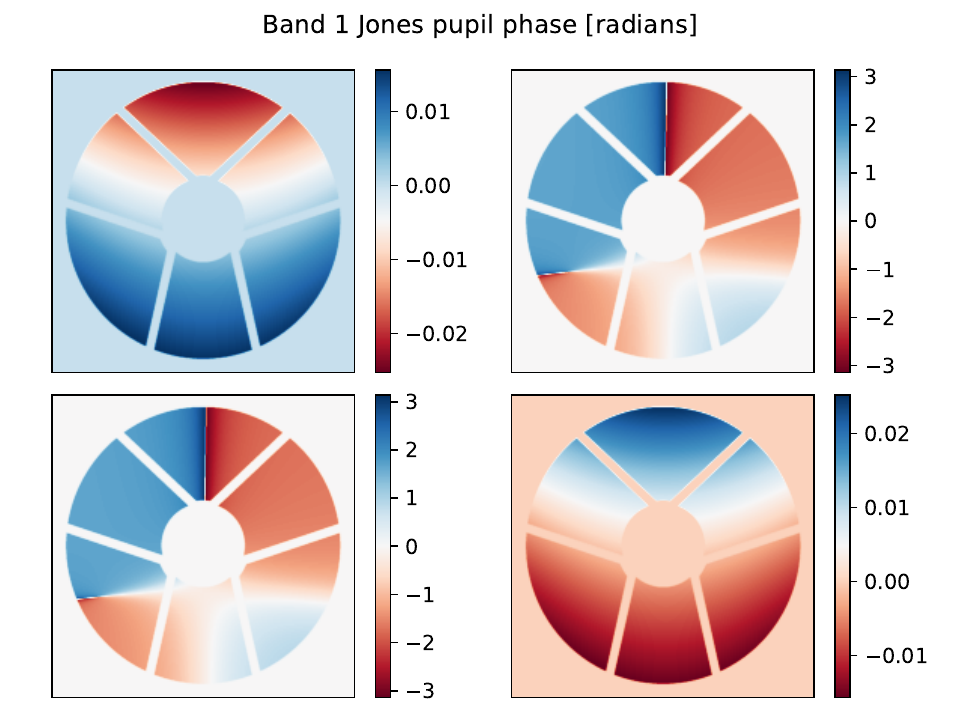}
      \end{tabular}
   \end{center}
   \caption[example] 
   { \label{fig:part_J_b1} The derotated and corrected Roman Jones pupils of band 1, where the phase is now in radians, in an XY $\rightarrow$ XY coordinate system. }
   \end{figure} 
Therefore, we confirm that the Roman Mueller matrix is not included in the FALCO Jones pupils, as expected from the removal of all piston terms. 
We do see similar structures in the phase and amplitude aberrations as we see for ground-based telescopes, where M1, M2, and M3 generate phase jumps in the off-diagonal terms.
The sign flip between band 1 and 4 in the phase aberrations is consistent with a flip in the sign of the retardance as found in Figure \ref{fig:ret_diat}.
    \begin{figure} [ht]
   \begin{center}
   \begin{tabular}{cc} 
   \includegraphics[width=0.47\linewidth]{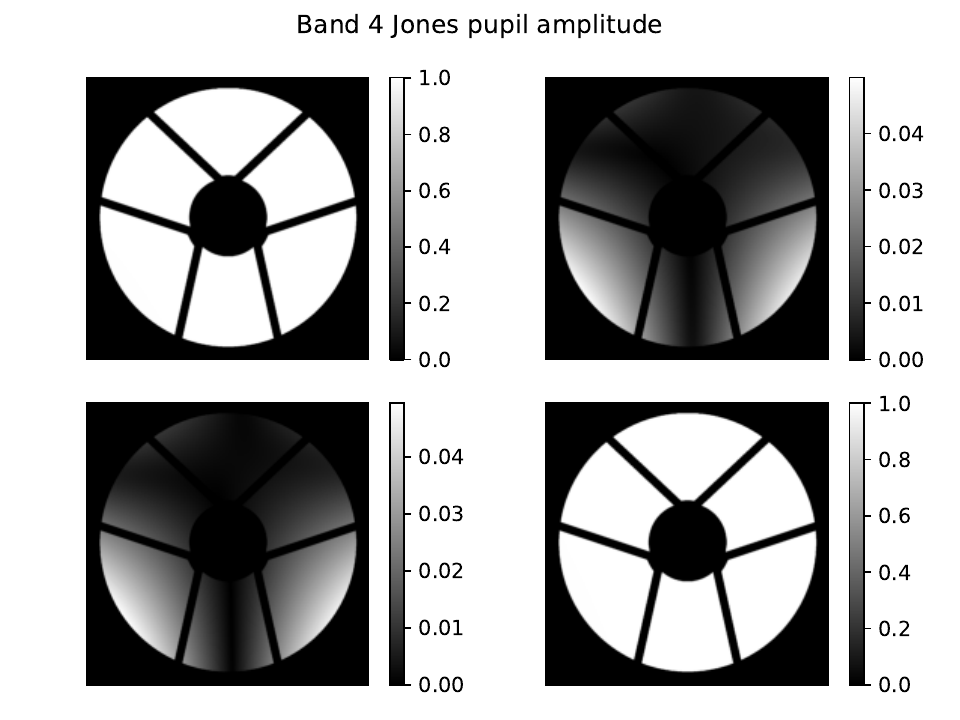} \includegraphics[width=0.47\linewidth]{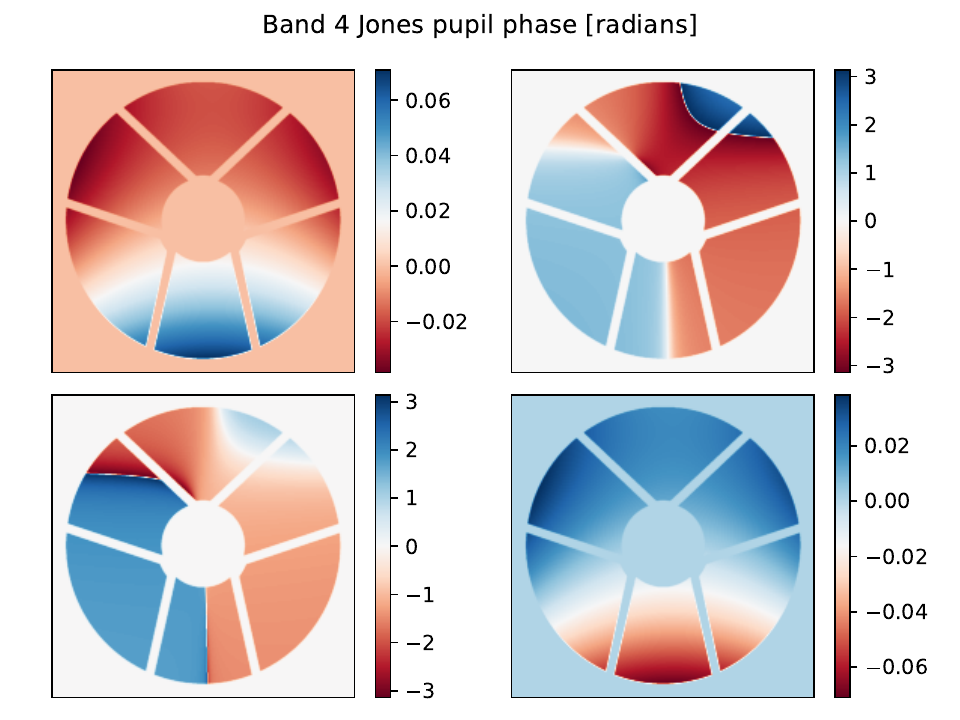}
      \end{tabular}
   \end{center}
   \caption[example] 
   { \label{fig:part_J_b4} The derotated and corrected Roman Jones pupils of band 4, where the phase is now in radians, in an XY $\rightarrow$ XY coordinate system}
   \end{figure}

\bibliography{RomanPol} 

\begin{thebibliography}{10}

\bibitem{milli2019optical}
Milli, J., Engler, N., Schmid, H., Olofsson, J., M{\'e}nard, F., Kral, Q.,
  Boccaletti, A., Th{\'e}bault, P., Choquet, E., Mouillet, D., et~al.,
  ``Optical polarised phase function of the hr 4796a dust ring,'' {\em
  Astronomy \& Astrophysics}~{\bf 626},  A54 (2019).

\bibitem{chen2020multiband}
Chen, C., Mazoyer, J., Poteet, C.~A., Ren, B., Duch{\^e}ne, G., Hom, J.,
  Arriaga, P., Millar-Blanchaer, M.~A., Arnold, J., Bailey, V.~P., et~al.,
  ``Multiband gpi imaging of the hr 4796a debris disk,'' {\em The Astrophysical
  Journal}~{\bf 898}(1),  55 (2020).

\bibitem{mclean2017polarimetric}
McLean, W., Stam, D., Bagnulo, S., Borisov, G., Devog{\`e}le, M., Cellino, A.,
  Rivet, J., Bendjoya, P., Vernet, D., Paolini, G., et~al., ``A polarimetric
  investigation of jupiter: Disk-resolved imaging polarimetry and
  spectropolarimetry,'' {\em Astronomy \& Astrophysics}~{\bf 601},  A142
  (2017).

\bibitem{stam2004using}
Stam, D., Hovenier, J., and Waters, L., ``Using polarimetry to detect and
  characterize jupiter-like extrasolar planets,'' {\em Astronomy \&
  Astrophysics}~{\bf 428}(2),  663--672 (2004).

\bibitem{rossi2021spectropolarimetry}
Rossi, L., Berzosa-Molina, J., Desert, J.-M., Fossati, L., Mu{\~n}oz, A.~G.,
  Haswell, C., Kabath, P., Kislyakova, K., Stam, D., and Vidotto, A.,
  ``Spectropolarimetry as a tool for understanding the diversity of planetary
  atmospheres,'' {\em Experimental Astronomy} ,  1--10 (2021).

\bibitem{quanz2011very}
Quanz, S.~P., Schmid, H.~M., Geissler, K., Meyer, M.~R., Henning, T., Brandner,
  W., and Wolf, S., ``Very large telescope/naco polarimetric differential
  imaging of hd100546—disk structure and dust grain properties between 10 and
  140 au,'' {\em The Astrophysical Journal}~{\bf 738}(1),  23 (2011).

\bibitem{stolker2016shadows}
Stolker, T., Dominik, C., Avenhaus, H., Min, M., De~Boer, J., Ginski, C.,
  Schmid, H.-M., Juh{\'a}sz, A., Bazzon, A., Waters, L., et~al., ``Shadows cast
  on the transition disk of hd 135344b-multiwavelength vlt/sphere polarimetric
  differential imaging,'' {\em Astronomy \& Astrophysics}~{\bf 595},  A113
  (2016).

\bibitem{vaughan2023chasing}
Vaughan, S.~R., Gebhard, T.~D., Bott, K., Casewell, S.~L., Cowan, N.~B.,
  Doelman, D.~S., Kenworthy, M., Mazoyer, J., Millar-Blanchaer, M.~A., Trees,
  V.~J., et~al., ``Chasing rainbows and ocean glints: Inner working angle
  constraints for the habitable worlds observatory,'' {\em Monthly Notices of
  the Royal Astronomical Society} ,  stad2127 (2023).

\bibitem{spergel2015wide}
Spergel, D., Gehrels, N., Baltay, C., Bennett, D., Breckinridge, J., Donahue,
  M., Dressler, A., Gaudi, B., Greene, T., Guyon, O., et~al., ``Wide-field
  infrarred survey telescope-astrophysics focused telescope assets wfirst-afta
  2015 report,'' {\em arXiv preprint arXiv:1503.03757}  (2015).

\bibitem{mennesson2018wfirst}
Mennesson, B., Debes, J., Douglas, E., Nemati, B., Stark, C., Kasdin, J.,
  Macintosh, B., Turnbull, M., Rizzo, M., Roberge, A., et~al., ``The wfirst
  coronagraph instrument: a major step in the exploration of sun-like planetary
  systems via direct imaging,'' in [{\em Space Telescopes and Instrumentation
  2018: Optical, Infrared, and Millimeter Wave}{\nolinebreak\hspace{0.1em}]},
  {\bf 10698},  779--787, SPIE (2018).

\bibitem{noecker2016coronagraph}
Noecker, M.~C., Zhao, F., Demers, R., Trauger, J., Guyon, O., and
  Jeremy~Kasdin, N., ``Coronagraph instrument for wfirst-afta,'' {\em Journal
  of Astronomical Telescopes, Instruments, and Systems}~{\bf 2}(1),
  011001--011001 (2016).

\bibitem{kasdin2020nancy}
Kasdin, N.~J., Bailey, V.~P., Mennesson, B., Zellem, R.~T., Ygouf, M., Rhodes,
  J., Luchik, T., Zhao, F., Riggs, A.~E., Seo, B.-J., et~al., ``The nancy grace
  roman space telescope coronagraph instrument (cgi) technology
  demonstration,'' in [{\em Space Telescopes and Instrumentation 2020: Optical,
  Infrared, and Millimeter Wave}{\nolinebreak\hspace{0.1em}]},   {\bf 11443},
  300--313, SPIE (2020).

\bibitem{krist2018wfirst}
Krist, J., Effinger, R., Kern, B., Mandic, M., McGuire, J., Moody, D.,
  Morrissey, P., Poberezhskiy, I., Riggs, A., Saini, N., et~al., ``Wfirst
  coronagraph flight performance modeling,'' in [{\em Space Telescopes and
  Instrumentation 2018: Optical, Infrared, and Millimeter
  Wave}{\nolinebreak\hspace{0.1em}]},   {\bf 10698},  788--810, SPIE (2018).

\bibitem{groff2021roman}
Groff, T.~D., Zimmerman, N.~T., Subedi, H.~B., Rizzo, M.~J., Titus, J., Lyons,
  J., Bell, D., Gaylin, S., Gao, G., Pasquale, B., et~al., ``Roman space
  telescope cgi: prism and polarizer characterization modes,'' in [{\em Space
  Telescopes and Instrumentation 2020: Optical, Infrared, and Millimeter
  Wave}{\nolinebreak\hspace{0.1em}]},   {\bf 11443},  640--648, SPIE (2021).

\bibitem{zellem2022nancy}
Zellem, R.~T., Nemati, B., Gonzalez, G., Ygouf, M., Bailey, V.~P., Cady, E.~J.,
  Colavita, M.~M., Hildebrandt, S.~R., Maier, E.~R., Mennesson, B., et~al.,
  ``Nancy grace roman space telescope coronagraph instrument observation
  calibration plan,'' in [{\em Space Telescopes and Instrumentation 2022:
  Optical, Infrared, and Millimeter Wave}{\nolinebreak\hspace{0.1em}]},   {\bf
  12180},  705--736, SPIE (2022).

\bibitem{maier2022flatfield}
Maier, E.~R., Zellem, R.~T., Colavita, M.~M., Mennesson, B., Nemati, B.,
  Bailey, V.~P., Cady, E.~J., Weisberg, C., Ryan, D., Belikov, R., et~al.,
  ``Flatfield calibrations with astrophysical sources for the nancy grace roman
  space telescope's coronagraph instrument,'' {\em arXiv preprint
  arXiv:2202.04815}  (2022).

\bibitem{snik2013astronomical}
Snik, F. and Keller, C.~U., ``Astronomical polarimetry: polarized views of
  stars and planets,'' {\em Planets, Stars and Stellar Systems. Volume 2:
  Astronomical Techniques, Software and Data} ,  175 (2013).

\bibitem{chipman1995mechanics}
Chipman, R.~A., ``Mechanics of polarization ray tracing,'' {\em Optical
  Engineering}~{\bf 34}(6),  1636--1645 (1995).

\bibitem{de2020polarimetric}
De~Boer, J., Langlois, M., van Holstein, R.~G., Girard, J.~H., Mouillet, D.,
  Vigan, A., Dohlen, K., Snik, F., Keller, C.~U., Ginski, C., et~al.,
  ``Polarimetric imaging mode of vlt/sphere/irdis-i. description, data
  reduction, and observing strategy,'' {\em Astronomy \& Astrophysics}~{\bf
  633},  A63 (2020).

\bibitem{van2020polarimetric}
van Holstein, R.~G., Girard, J.~H., De~Boer, J., Snik, F., Milli, J., Stam, D.,
  Ginski, C., Mouillet, D., Wahhaj, Z., Schmid, H.~M., et~al., ``Polarimetric
  imaging mode of vlt/sphere/irdis-ii. characterization and correction of
  instrumental polarization effects,'' {\em Astronomy \& Astrophysics}~{\bf
  633},  A64 (2020).

\bibitem{wiktorowicz2014gemini}
Wiktorowicz, S.~J., Millar-Blanchaer, M., Perrin, M.~D., Graham, J.~R.,
  Fitzgerald, M.~P., Maire, J., Ingraham, P., Savransky, D., Macintosh, B.~A.,
  Thomas, S.~J., et~al., ``Gemini planet imager observational calibrations vii:
  on-sky polarimetric performance of the gemini planet imager,'' in [{\em
  Ground-based and Airborne Instrumentation for Astronomy
  V}{\nolinebreak\hspace{0.1em}]},   {\bf 9147},  2574--2584, SPIE (2014).

\bibitem{millar2016gpi}
Millar-Blanchaer, M.~A., Perrin, M.~D., Hung, L.-W., Fitzgerald, M.~P., Wang,
  J.~J., Chilcote, J., Graham, J.~R., Bruzzone, S., and Kalas, P.~G., ``Gpi
  observational calibrations xiv: polarimetric contrasts and new data reduction
  techniques,'' in [{\em Ground-based and Airborne Instrumentation for
  Astronomy VI}{\nolinebreak\hspace{0.1em}]},   {\bf 9908},  993--1009, SPIE
  (2016).

\bibitem{van2020calibration}
van Holstein, R.~G., Bos, S.~P., Ruigrok, J., Lozi, J., Guyon, O., Norris, B.,
  Snik, F., Chilcote, J., Currie, T., Groff, T.~D., et~al., ``Calibration of
  the instrumental polarization effects of scexao-charis’spectropolarimetric
  mode,'' in [{\em Ground-based and Airborne Instrumentation for Astronomy
  VIII}{\nolinebreak\hspace{0.1em}]},   {\bf 11447},  1113--1126, SPIE (2020).

\bibitem{gj2021full}
GJ't~Hart, J., van Holstein, R.~G., Bos, S.~P., Ruigrok, J., Snik, F., Lozi,
  J., Guyon, O., Kudo, T., Zhang, J., Jovanovic, N., et~al., ``Full
  characterization of the instrumental polarization effects of the
  spectropolarimetric mode of scexao/charis,'' in [{\em Polarization Science
  and Remote Sensing X}{\nolinebreak\hspace{0.1em}]},   {\bf 11833},  148--174,
  SPIE (2021).

\bibitem{chipman1987polarization}
Chipman, R.~A.,  [{\em Polarization aberrations (thin
  films)}{\nolinebreak\hspace{0.1em}]}, The University of Arizona (1987).

\bibitem{mcguire1987polarization}
McGuire~Jr, J.~P. and Chipman, R.~A., ``Polarization aberrations in optical
  systems,'' in [{\em Current Developments in Optical Engineering
  II}{\nolinebreak\hspace{0.1em}]},   {\bf 818},  240--257, SPIE (1987).

\bibitem{chipman2015polarization}
Chipman, R.~A., Lam, W. S.~T., and Breckinridge, J., ``Polarization aberration
  in astronomical telescopes,'' in [{\em Polarization Science and Remote
  Sensing VII}{\nolinebreak\hspace{0.1em}]},   {\bf 9613},  124--134, SPIE
  (2015).

\bibitem{van_Holstein_2023}
van Holstein, R., Keller, C., Snik, F., and Bos, S., ``Polarization-dependent
  beam shifts upon metallic reflection in high-contrast imagers and
  telescopes,'' {\em Astronomy \& Astrophysics}  (jul 2023).

\bibitem{breckinridge2015polarization}
Breckinridge, J.~B., Lam, W. S.~T., and Chipman, R.~A., ``Polarization
  aberrations in astronomical telescopes: the point spread function,'' {\em
  Publications of the Astronomical Society of the Pacific}~{\bf 127}(951),  445
  (2015).

\bibitem{anche2018analysis}
Anche, R.~M., Sen, A.~K., Anupama, G.~C., Sankarasubramanian, K., and Skidmore,
  W., ``Analysis of polarization introduced due to the telescope optics of the
  thirty meter telescope,'' {\em Journal of Astronomical Telescopes,
  Instruments, and Systems}~{\bf 4}(1),  018003--018003 (2018).

\bibitem{anche2023estimation}
Anche, R.~M., Williams, G., Packham, C., Ashcraft, J., and Douglas, E.~S.,
  ``Estimation of polarization aberrations and their effect on the
  coronagraphic performance for future space telescopes,'' in [{\em Techniques
  and Instrumentation for Detection of Exoplanets
  XI}{\nolinebreak\hspace{0.1em}]},   {\bf 12680-28}, SPIE (2023).

\bibitem{anche2023polarimetric}
Anche, R.~M., Haffert, S.~Y., Ashcraft, J., Van~Gorkom, K., Derby, K., Douglas,
  E.~S., and Millar-Blanchaer, M.~A., ``Polarimetric modeling and assessment of
  science cases for giant magellan telescope-polarimeter (gmt-pol),'' in [{\em
  Polarization Science and Remote Sensing XI}{\nolinebreak\hspace{0.1em}]},
  {\bf 12690-20}, SPIE (2023).

\bibitem{anche2023polarization}
Anche, R.~M., Ashcraft, J.~N., Haffert, S.~Y., Millar-Blanchaer, M.~A.,
  Douglas, E.~S., Snik, F., Williams, G., van Holstein, R.~G., Doelman, D.,
  Van~Gorkom, K., et~al., ``Polarization aberrations in next-generation giant
  segmented mirror telescopes (gsmts) i. effect on the coronagraphic
  performance,'' {\em arXiv preprint arXiv:2304.02079}  (2023).

\bibitem{schmid2018sphere}
Schmid, H.~M., Bazzon, A., Roelfsema, R., Mouillet, D., Milli, J., Menard, F.,
  Gisler, D., Hunziker, S., Pragt, J., Dominik, C., et~al., ``Sphere/zimpol
  high resolution polarimetric imager-i. system overview, psf parameters,
  coronagraphy, and polarimetry,'' {\em Astronomy \& Astrophysics}~{\bf 619},
  A9 (2018).

\bibitem{millar2022polarization}
Millar-Blanchaer, M.~A., Anche, R.~M., Nguyen, M.~M., and Maire, J., ``The
  polarization aberrations of the gemini telescope as seen by the gemini planet
  imager,'' in [{\em Ground-based and Airborne Instrumentation for Astronomy
  IX}{\nolinebreak\hspace{0.1em}]},   {\bf 12184},  1278--1288, SPIE (2022).

\bibitem{sanchez1992instrumental}
Sanchez~Almeida, J. and Martinez~Pillet, V., ``Instrumental polarization in the
  focal plane of telescopes,'' {\em Astronomy and Astrophysics (ISSN
  0004-6361), vol. 260, no. 1-2, p. 543-555.}~{\bf 260},  543--555 (1992).

\bibitem{sanchez1994instrumental}
S{\'a}nchez~Almeida, J., ``Instrumental polarization in the focal plane of
  telescopes. 2: effects induced by seeing,'' {\em Astronomy and Astrophysics
  (ISSN 0004-6361), vol. 292, no. 2, p. 713-721}~{\bf 292},  713--721 (1994).

\bibitem{riggs2018fast}
Riggs, A.~E., Ruane, G., Sidick, E., Coker, C., Kern, B.~D., and Shaklan,
  S.~B., ``Fast linearized coronagraph optimizer (falco) i: a software toolbox
  for rapid coronagraphic design and wavefront correction,'' in [{\em Space
  Telescopes and Instrumentation 2018: Optical, Infrared, and Millimeter
  Wave}{\nolinebreak\hspace{0.1em}]},   {\bf 10698},  878--888, SPIE (2018).

\bibitem{perrin2015polarimetry}
Perrin, M.~D., Duchene, G., Millar-Blanchaer, M., Fitzgerald, M.~P., Graham,
  J.~R., Wiktorowicz, S.~J., Kalas, P.~G., Macintosh, B., Bauman, B., Cardwell,
  A., et~al., ``Polarimetry with the gemini planet imager: methods, performance
  at first light, and the circumstellar ring around hr 4796a,'' {\em The
  Astrophysical Journal}~{\bf 799}(2),  182 (2015).

\bibitem{van2021survey}
van Holstein, R.~G., Stolker, T., Jensen-Clem, R., Ginski, C., Milli, J.,
  De~Boer, J., Girard, J., Wahhaj, Z., Bohn, A., Millar-Blanchaer, M., et~al.,
  ``A survey of the linear polarization of directly imaged exoplanets and brown
  dwarf companions with sphere-irdis-first polarimetric detections revealing
  disks around dh tau b and gsc 6214-210 b,'' {\em Astronomy \&
  Astrophysics}~{\bf 647},  A21 (2021).

\bibitem{mendillo2021dual}
Mendillo, C.~B., Hewawasam, K., Martel, J., Cook, T.~A., Chakrabarti, S., Snik,
  F., and Doelman, D., ``Dual-polarization electric field conjugation and
  applications for vector vortex coronagraphs,'' in [{\em Techniques and
  Instrumentation for Detection of Exoplanets X}{\nolinebreak\hspace{0.1em}]},
   {\bf 11823},  586--594, SPIE (2021).

\end{thebibliography}
\bibliographystyle{spiebib} 

\end{document}